 \documentclass[preprint,flushrt,tighten]{aastex}
 \usepackage{apjfonts,emulateapj5}

\newcommand{\euve}{\textit{EUVE}}
\newcommand{\fnref}[1]{\mbox{$^{\ref{#1}}$}}
\newcommand{\kms}{km~s$^{-1}$}
\newcommand{\cps}{ct~s$^{-1}$}
\newcommand{\ergps}{erg~s$^{-1}$}
\newcommand{\lxlbol}{$L_\mathrm{X}/L_\mathrm{bol}$}

\slugcomment{To Appear in The Astrophysical Journal}

\received{March 7, 2000}
\revised{April 26, 2000}
\accepted{April 28, 2000}

\shorttitle{EUV FLARE ACTIVITY IN LATE-TYPE STARS}
\shortauthors{AUDARD ET AL.}

\begin{document}


\title{EUV FLARE ACTIVITY IN LATE-TYPE STARS}


\author{Marc Audard\altaffilmark{1,2}, Manuel G\"udel\altaffilmark{1,2}, Jeremy J.
Drake\altaffilmark{3}, and Vinay L. Kashyap\altaffilmark{3}}

\altaffiltext{1}{Laboratory for Astrophysics, Paul Scherrer Institute, W\"urenlingen \& Villigen, 5232 Villigen PSI,
Switzerland}
\altaffiltext{2}{Mailing address: Institute of Astronomy, ETH Zentrum, 8092 Z\"urich,
Switzerland; \texttt{audard@astro.phys.ethz.ch},
\texttt{guedel@astro.phys.ethz.ch}}
\altaffiltext{3}{Harvard-Smithsonian Center for Astrophysics, Cambridge, MA 02138;
\texttt{jdrake@cfa.harvard.edu}, \texttt{vkashyap@cfa.harvard.edu}}


\begin{abstract}
\textit{Extreme Ultraviolet Explorer} Deep Survey observations of cool stars (spectral type F to M) have been used to
investigate the distribution of coronal flare rates in energy and its relation to activity indicators 
and rotation parameters. Cumulative and differential flare rate
distributions were constructed and fitted with
different methods. Power laws are found to approximately describe the distributions. A trend toward
flatter distributions for later-type stars is suggested in our sample.
Assuming that the power laws continue below the detection
limit, we have estimated that the superposition of flares with radiated
energies of about $10^{29}-10^{31}$~ergs could explain
the observed radiative power loss of these coronae, while the detected
flares are contributing only $\approx 10$~\%.
While the power-law index is not correlated with rotation parameters
(rotation period, projected rotational velocity, Rossby number) and only
marginally with the X-ray luminosity, the flare occurrence rate is correlated
with all of them. The occurrence rate of flares with energies larger than
$10^{32}$~ergs is found to be proportional to the average total stellar X-ray
luminosity. Thus, energetic flares occur more often in X-ray
bright stars than in X-ray faint stars. The normalized occurrence rate of flares with energies larger than
$10^{32}$~ergs increases with increasing \lxlbol\ and stays constant
for saturated stars. A similar saturation is found below a critical Rossby number.
The findings are discussed in terms of simple statistical flare models in an attempt to explain the
previously observed trend for higher average coronal temperatures in more active stars. It is concluded
that flares can contribute a significant amount of energy to
coronal heating in active stars.
\end{abstract}


\section{INTRODUCTION}

Stellar activity of ``normal'' stars has been
explored extensively in the X-ray regime for more than two decades \citep[e.g.,][]{vaiana81,linsky85}. The
chromospheres and coronae of some late-F to M main-sequence stars have been found to show enhanced
magnetic activity. The latter is underlined by enhanced activity indicators such as the X-ray luminosity
$L_\mathrm{X}$, its ratio to the bolometric luminosity \lxlbol, the presence of flares in
the optical $U\!$ band (and in other wavelength regions), flux variations in chromospheric lines,
spots on the stellar surface, etc. A dynamo mechanism is thought to be the
primary cause for stellar and solar activity as suggested by the empirical relation
between rotation and activity \citep[for a recent review, see][]{simon2000}.
Rotation parameters (such as the rotation period, the Rossby number,
or the angular velocity) are therefore prime parameters that determine the
stellar activity level.

Stellar rotation was proposed to determine the level of activity of solar-type
stars by \citet{kraft67}. Quantitative relationships between activity indicators
and rotation parameters \citep[e.g.,][]{skum72,pall81,walt82,noy84,rand96}
provide information on the physical origin of stellar activity. In a study of X-ray 
emission from stars, \citet{pall81} found that X-ray luminosities of late-type
stars are dependent on the projected rotational
velocity but are independent of bolometric luminosity. 
\citet{walt81a} and \citet{walt81b} presented observational evidence that RS
CVn systems and G-type stars show a quiescent \lxlbol\ ratio proportional
to their angular velocity. However, in the same series of papers, \citet{walt82} proposed that the
simplistic view of a power-law dependence should be replaced by either a broken power
law or by an exponential dependence of \lxlbol\ on
the angular velocity. This led to the concept of saturation of stellar
activity \citep{vilhu84,vilhu87} at high rotation rate. \citet{noy84} suggested that the Rossby
number $R_0$ (defined as the ratio $P/\tau_\mathrm{c}$ of the rotation period $P$ and the
convective turnover time $\tau_\mathrm{c}$) mainly determines the surface magnetic activity
in lower main-sequence stars. \citet{stepien93} showed that, for main-sequence
late-type stars, $R_0$ correlates with activity indicators better than $P$ does.

Flares are direct evidence of magnetic activity in stellar atmospheres. They are
also in the center of the debate on the origin of coronal heating. Although
several possible heating mechanisms have been identified
\citep[e.g.,][]{ionson85,narain90,zirker93,haisch96}, there is increasing
evidence that flares act as heating agents of the outer atmospheric layers of
stars. A correlation between the apparently
non-flaring (``quiescent'') coronal X-ray luminosity $L_\mathrm{X}$ and the stellar
time-averaged $U$-band flare luminosity \citep{doyle85,skum85} suggests
that flares can release a sufficient amount of energy to produce the subsequently observed quiescent
coronal emission. \citet{robinson95,robinson99} found evidence for numerous transition
region (TR) flares in CN Leo and YZ CMi. Further evidence for dynamic
heating has been found in broadened TR emission line profiles
\citep{linsky94,wood96} that were interpreted in terms of a large number of
explosive events.

Stellar X-ray and EUV flares have been found to be distributed in energy according to
a power law \citep*[see][]{collura88,aud99,osten99}, similar to 
solar flares \citep*[e.g,][]{crosby93}. One finds, for the
flare rate $dN$ within the energy interval [$E,E+dE$],
\begin{equation}
\frac{dN}{dE} = k_1 E^{-\alpha}.\label{k1def}
\end{equation}
The cumulative distribution for $\alpha > 1$ is defined by
\begin{mathletters}
\begin{eqnarray}
N(>E)	& = & \int_{E}^{\infty}\frac{dN}{dE^\prime}dE^\prime,\\
	& = & k_2 E^{-\alpha+1},  \label{k2def}
\end{eqnarray}
\end{mathletters}
where the normalization factors $k_1$ and $k_2 = k_1 / \left( \alpha - 1
\right)$ are constants. For $\alpha > 2$, an extrapolation to flare energies
below the instrumental detection limit could be sufficient to produce a radiated
power equivalent to the X-ray luminosity $L_\mathrm{X}$ of the quiescent corona, 
\begin{mathletters}
\begin{eqnarray}
L_\mathrm{X} & = & \int_{E_\mathrm{min}}^{E_{\mathrm{max}}}\frac{dN}{dE} E ~dE,\\
    & = & k_2\frac{\alpha-1}{\alpha-2} \left(
    E_\mathrm{min}^{2-\alpha}-E_{\mathrm{max}}^{2-\alpha} \right),
\label{power}
\end{eqnarray}
\end{mathletters}
\noindent where $E_\mathrm{max}$ is the energy of the most energetic relevant
flare. Small values of the minimum flare energy $E_\mathrm{min}$ then can lead
to an arbitrarily
large radiated power. Thus, as pointed out by \citet{hudson91}, it is crucial
to investigate whether the solar (and stellar) flare rate distributions in
energy steepen at lower energies. \citet{parker88} suggested that the heating of the
quiescent solar corona could be explained by ``microflares''. In the
solar context, several estimates of the power-law index have recently been given.
For ``normal'' flares, $\alpha \approx 1.5 - 1.8$ \citep{crosby93};
it takes different values for smaller flare energies, from $\alpha \approx 1.6$
for small active-region transient brightenings \citep{shimizu95} to $\alpha =
2.3 - 2.6$ for small events in the quiet solar corona 
\citep{krucker98}, and other ``intermediate'' values
\citep*[e.g.,][]{porter95,aschwanden00,parnell00}. Stellar
studies on flare occurrence distributions in X-rays are rare, probably due to the paucity of
stellar flare statistics. Using \textit{EXOSAT} data, \citet{collura88} found
a power-law index $\alpha$ of 1.52 for soft X-ray flares on M dwarfs. 
\citet{osten99} reported $\alpha = 1.6$ for flares in RS
CVn systems observed with \euve. On the other hand,
\citet{aud99} found a power-law index $\alpha \approx
2.2 \pm 0.2$ for two young active solar analogs.

In this paper, we present a follow-up study of \citet{aud99} for
\euve\ observations of active late-type main-sequence stars. Together with an
investigation of coronal heating by flares, a more general
picture of the relation between flares and stellar activity in general is developed.
Section~\ref{sectdata} presents the method for the data selection and
reduction, section~\ref{constrdistr} explains the construction of cumulative
and differential flare rate distributions in energy. Section~\ref{sectfits}
provides the methodology for fitting the distributions, while
section~\ref{sectexpl}
explores quantitatively the correlations between various physical parameters.
Finally, section~\ref{sectdisc} gives a discussion of the results, together with
conclusions.

\section{DATA SELECTION AND REDUCTION}\label{sectdata}
We use data from the \textit{Extreme Ultraviolet Explorer} \citep[\euve, e.g.][]{malina91} to study
the contribution of flares to the observable EUV and X-ray emission from stellar coronae. In order to
identify a sufficient number of flares in the \euve\ Deep Survey (DS) light curves, data sets with
more than 5 days of monitoring, or with a significant number of flares (more
than ten flares identified by eye) were selected. Active coronal sources were our
prime choice, as these stars often show several distinct stochastic events. We
have focused our analysis on young, active stars that do not display rotationally modulated 
light curves and that can be considered as single X-ray sources. Some stars in
the sample are detected or known binary systems, in which
only one of the components is believed to contribute significantly to
the EUV and X-ray emitted radiation. If a star was observed several times
(more than a few days apart), we considered the different data sets 
as originating from different coronal sources, as the activity level of these
stars is usually not identical at two different epochs. We carefully 
checked the DS data and rejected data that presented evident problems, such as 
``ghost'' images in the DS remapped event files, or incursions into the DS ``dead 
spot''. The final list contains 12 stellar sources (1 F-type, 4 G-type, 
2 K-type, and 5 M-type coronal sources). We do not claim our sample to be
complete in any sense. However, this sample is representative of the 
content of the magnetically active cool main-sequence stellar population in the \euve\ archive.
Table~\ref{list} gives the name of the stellar
source (Col.~1), its spectral type (Col.~2), its distance $d$ in parsecs from
Hipparcos (\citealt*{perry97}; except for AD Leo and CN Leo which are from
\citealt*{gliese91}; Col.~3), the rotation period $P$ in days 
together with its reference (Cols.~4 \& 5), the projected rotational
velocity\footnote{For 47 Cas, the X-ray emitter
is the probable, optically hidden G0--5~V companion \citep{guedel00}, therefore we have set $\bv = 0.62$. 
We have estimated the equatorial velocity from the rotation period of the X-ray
bright source, from the bolometric luminosity and an effective temperature of 
$T_\mathrm{eff} \approx 5900$~K.\label{47casdef}} and its
reference (Cols.~6 \& 7), the color index \bv\fnref{47casdef} and the visual magnitude $V$ from
Hipparcos (\citealt*{perry97}; except for AD Leo and CN Leo for which data were
retrieved from Simbad; Cols.~8 \& 9), the mean DS count rate (Col.~10), the derived (see below) 
EUV+X-ray (hereafter ``coronal'') luminosity in the 0.01--10~keV energy range (Col.~11), and the \euve\ 
observing window (Col.~12).

\placetable{list}

We have made extensive use of the data from the 
\anchor{http://archive.stsci.edu/cgi-bin/euve}{\euve\ archive} located at the
Multimission Archive at the Space Telescope Science Institute
(STScI)\footnote{IRAF is distributed by the National Optical Astronomy Observatories
(NOAO). STScI and NOAO are operated by the Association of Universities for
Research in Astronomy, Inc.\label{aura}}. DS Remapped Archive QPOE files were rebuilt using the
\texttt{euv1.8} package within IRAF\fnref{aura}. Light curves (Fig.~\ref{lcsigdef})
were created using a DS background region ten times larger than the 
source region area. Event lists were extracted from the source region
for further analysis. Thanks to the sufficiently large count rates of our sources, the contribution of the
DS background was very low, and could be neglected (after a check for its constancy). Also, 
with our analysis method, flare-only count rates (count rates above the 
``quiescent'' level) were required; therefore the small contribution of the background 
was eliminated in any case. Event and Good Time Interval
(GTI) files were then read and processed with a flare identification code.
\citet{aud99} explain in detail the procedure applied to identify flares in the DS
event files. In brief, the method, adapted from \citet{robinson95}, performs a
statistical identification of flares. It assigns occurrence probabilities to
light-curve bins. Note that several time bin lengths and time origins for the binning
are used so that the identification of flares is not dependent on the choice of
these parameters. Note also that,
due to gaps between GTIs, the effective exposure of a bin had to be taken
into account. We refer to \citet{robinson95} and \citet{aud99} for more details.

Figure~\ref{lcsigdef} (upper panels) shows background-subtracted \euve\
DS light curves\footnote{With task \texttt{qpbin} of \texttt{euv1.8}, 
the last bin of a light-curve plot is generally omitted \citetext{D.~J. Christian~1999,
priv.~comm.}} for some data sets. Only for plotting purposes, the data have been binned 
to one bin per orbit ($P_\mathrm{orb} = 96$~min). Note that for our data analysis, we have not
restricted ourselves to the above bin size: bin durations from 1/5 to twice the orbital period have
been used (see Fig.~\ref{lcsigdef}). The lower panels show the
corresponding ``significance plots'', which give the probability for the presence of 
quiescent bins as a function of time (x-axis) and bin size (y-axis). 
The flare significance increases from light gray to black. Physical
parameters (start time, end time and total duration) were determined from 
smoothed high-resolution light curves, 
using a Gaussian fit to the flares above a smooth lower envelope characterizing the quiescent 
contribution.  Thus the start and end times of a flare were defined as the times separated from 
the maximum  by 2~$\sigma$, where $\sigma$ is the standard deviation of the Gaussian
function. Within this interval, we calculated a mean flare count rate above
the quiescent emission by subtracting the mean background levels just before and
after the flare, and multiplied it by the flare total duration to derive 
the total number of flare counts $C$. We then used a constant count-to-energy
conversion factor ($f = 1.06 \times 10^{27} \; \mathrm{ergs\;ct^{-1}\;pc^{-2}}$) 
together with the source distance $d$ to derive the total energy $E$ radiated in the
EUV and X-rays,
\begin{equation}
E  =  C \times f \times \left( 4  \pi  d^2 \right).
\end{equation}
We derived the conversion factor $f$ from mean DS count rates of archival \euve\ cool-star 
data sets and published X-ray luminosities
(\citealt*{pall88,vdo88,pall90,demp93a,demp93b,sch95,mons96,demp97,tagl97}; \citealt{aud99,hun99,scio99}).
We corrected the published X-ray luminosities to the new, Hipparcos-derived distances 
\citep{perry97}. Then, using a typical model for young active stars 
(2-temperature collisional ionization equilibrium \textsc{MEKAL} model [$T_{1} =
0.6$~keV, $T_{2} = 2.0$~keV] with the iron abundance $\mathrm{Fe} = 0.3$ 
times the solar photospheric value), we estimated factors to apply to
the published luminosities in order to convert them to 0.01--10~keV luminosities. We
finally defined the conversion factor $f$ as the mean ratio between the observed
fluxes $L_\mathrm{X}/(4 \pi d^2)$ and the mean DS count rates $\mu$.
Note that for our targets, the corona radiates mostly in the X-ray band ($\approx 0.1 - 5$~keV) 
rather than in the EUV band.

\placefigure{lcsigdef}

\section{FLARE OCCURRENCE RATE DISTRIBUTIONS} \label{constrdistr}
Cumulative flare rate distributions in energy were constructed for each source
(Fig.~\ref{cumdis}). Similarly to \citet{aud99}, we applied a correction
to the effective rate of identified flares. For each
cumulative distribution, the flare rate at the energy of the second-largest
flare was corrected by a factor $D_\mathrm{total}/(D_\mathrm{total}-D_\mathrm{largest})$, 
where $D_\mathrm{total}$ is the total observing time span, and 
$D_\mathrm{largest}$ is the total duration of the 
largest flare. Analogously, the correction for the third-largest flare took into 
account the total durations of the largest and second-largest flares, and so on.
This correction was necessary since usually more than 50~\% of the
DS light curves were occupied by identified flares, i.e. it is common for flares
to overlap in time.

For each cumulative distribution, we
have two series of parameters, namely the flare energy ($E_i$) and the flare
rate at this energy ($s_i$), with the indices
running from 0 (largest-energy flare, $s_0=1/D_\mathrm{total}$ by definition) to $M$
(lowest-energy flare). The cumulative occurrence rates $N(>E_i)$, i.e., the rate of
flares per day with energies exceeding $E_i$, were defined as
\begin{mathletters}
\begin{eqnarray}
N(>E_0) & = & \frac{1}{D_\mathrm{total}} = s_0,\\
N(>E_i) & = & N(>E_{i - 1}) + s_{i}, \quad i = 1, \ldots, M.
\end{eqnarray}
\end{mathletters}
We then constructed differential distributions. 
For the energy interval [$E_i,E_{i+1}$], we defined differential
flare occurrence rates ($n_i$) as
\begin{equation}
n_i = \frac{s_i}{E_{i}-E_{i+1}}, \quad i = 0, \ldots, M-1.
\end{equation}
We calculated uncertainties for the flare occurrence rates per unit
energy; we assumed that each number of flares
($s^\prime_i = s_i \times D_\mathrm{total}$) with energy $E_{i}$
has an uncertainty $\Delta s^\prime_i$ estimated from a Poisson distribution.
This approximation accounts for the larger relative uncertainty ($\Delta s^\prime_i/s^\prime_i$) 
of the number of detected flares at higher flare energies than at lower flare energies. 
However, instead of setting $\Delta s^\prime_i = \sqrt{s^\prime_i}$, we have
used the approximations proposed by
\citet{geh86}, who showed that for small $k$ following a Poisson distribution,
the 1~$\sigma$ upper error bar can be approximated with 
$1 + \left( k + 3/4 \right) ^{1/2}$, while the the 1~$\sigma$ lower
error bar is still approximated by the usual definition $\sqrt{k}$. 
Therefore, we have defined $\Delta s^\prime_i$ as the geometrical mean of the upper and 
lower error bars:
\begin{equation}
\Delta s^\prime_i = \left\{ \sqrt{s^\prime_i} \times \left( 1 + 
\sqrt{s^\prime_i + \case{3}{4}} \right ) \right\}^{1/2}.
\end{equation}
It follows that each differential occurrence rate $n_i$ has an uncertainty $\Delta n_i$ equal to
\begin{equation}
\Delta n_i = \frac{\Delta s^\prime_i/D_\mathrm{total}}{E_{i}-E_{i+1}}.
\end{equation}

\placefigure{cumdis}

\section{FITS TO THE DISTRIBUTIONS}\label{sectfits}
\subsection{Cumulative Distributions}
The power-law fitting procedure to the cumulative flare occurrence rate distributions is
adapted from \citet*{craw70}. This method is based on a maximum-likelihood
(ML) derivation of the best-fit power-law index $\alpha$. The best-fit
normalization factors $k_2$ (see eq.~[\ref{k2def}]) were then computed; using equation~(\ref{power}), the minimum flare
energy $E_\mathrm{min}$ required for the power law to explain the mean observed radiative 
energy loss ($L_\mathrm{X}$) was calculated, assuming that the cumulative flare occurrence 
rate distribution in radiated energy follow the same power law below the flare energy 
detection limit:
\begin{equation}
E_\mathrm{min} = \left\{ \frac{L_\mathrm{X}}{k_2} \left( \frac{\alpha - 2}{\alpha - 1} \right) + E_{\mathrm{max}}^{2 -
\alpha} \right\}^{1/(2 - \alpha)},
\label{emin}
\end{equation}
where the coronal luminosity $L_\mathrm{X}$ was estimated from $\mu$, the mean DS count rate,
\begin{equation}
L_\mathrm{X} = \mu \times f \times \left( 4 \pi d^{2} \right).
\label{lx}
\end{equation}
Columns 2 and 3  of Table~\ref{tabresult} give the power-law indices $\alpha$ of the 
cumulative distributions, and the corresponding minimum flare energies $E_\mathrm{min}$. The best-fit
power-law indices $\alpha$ derived from simple linear fits ($\chi^2$ method) in
the $\log N(>E) - \log E$
plane have been added for comparison in Column 4.

We find a possible trend for decreasing power-law indices with increasing color indices; 
sources of spectral type F or G tend to show indices above the 
critical value of 2, although $\alpha < 2$ is acceptable within the 1~$\sigma$ confidence range
(except for the F star HD 2726). On the other hand, K and M stars show 
various power-law indices, with $\alpha > 2$ being usually marginally
acceptable, although some individual sources show best-fit values 
above 2. The minimum flare energies $E_\mathrm{min}$ can be associated with relatively small
\emph{stellar} flare energies. In the solar context, however, they correspond to
medium-to-large flares ($E \approx 10^{29}-10^{31}$~ergs).

\placetable{tabresult}

\subsection{Differential Distributions}
Our cumulative distributions do not account for
uncertainties in the flare occurrence rates. Differential distributions allow us to
avoid this effect, and they include uncertainties in a natural way (see
section~\ref{constrdistr}). Therefore, we propose to use this different approach
in order to compare the results.
The differential distributions were transformed into FITS files
and were read into the XSPEC 10.00 software \citep{xspec}. Due to the characteristics of XSPEC,
energy bins were created in which each bin $\delta E_i$ is defined as the interval
between two consecutive flare energies (namely $\delta E_i = [ E_{i+1}, E_{i}]$, 
$i = 0,\ldots,M-1$). Finally, a power-law fit (implemented in the software,
using the $\chi^2$ minimization method) was performed for each 
distribution. To estimate the uncertainties derived for the index $\alpha$,
confidence ranges for a single parameter were calculated by varying the index
$\alpha$ and fitting the distribution until the deviation of $\chi^2$ from its 
best-fit value reached $\delta\chi^2= 1.00$. The power-law indices and their
corresponding confidence ranges can be found in Col.~5 (Table~\ref{tabresult}). 
Note that the small ``signal-to-noise'' of the distributions did not allow us
to better determine the confidence ranges. Relative uncertainties of the
$n_i$ values were usually larger than 50~\%, reaching about 150~\% at most. We
also note that the
values of $\alpha$ derived from fits to the cumulative distributions are similar to those
derived from fits to differential distributions;
confidence ranges for the second method are, however, larger and originate from
the inclusion of uncertainties in the flare rates, together with the small
number of detected flares. A strong support for this statement comes from the confidence range derived for $\kappa$~Cet 1994.
Only four energy bins were used, leading to large confidence ranges and an 
unconstrained upper limit for $\alpha$. 

\subsubsection{Combined Data Sets}\label{sptyp}
A possible dependence of $\alpha$ with the stellar spectral type has been mentioned
above. To test this trend further and also to obtain tighter results for our
power-law fits, we have performed \emph{simultaneous} fits to the 
differential distributions within XSPEC. In brief, all data sets belonging to a
spectral type (we combined the only F star with the G-type sources) were fitted
simultaneously with power laws of identical index $\alpha$ and one normalization
factor for each data set. Note that, with this procedure, the confidence range for $\alpha$
is better determined than in the case of individual distribution fits. The implicit
hypothesis of this procedure assumes that the coronae of stellar sources within
a given spectral class behave similarly, without any influence by
age, rotation period, projected stellar velocity, etc. Columns 6 \& 7 of Table~\ref{tabresult}
show the result of the simultaneous fits, together with 68.3~\% confidence
ranges for a single parameter. Again, we find a trend for lower indices at
later spectral types, although the significance is marginal at best.

\section{CORRELATIONS WITH PHYSICAL PARAMETERS}\label{sectexpl}
We have explored correlations of the best-fit power-law indices $\alpha$ and 
occurrence rates of flares showing energies larger than a typical energy observed
in our data ($E_\mathrm{c} = 10^{32}$~ergs) with rotation parameters (rotation period $P$, projected
rotational velocity $v \sin i$, Rossby number $R_0$) and activity indicators
(coronal luminosity $L_\mathrm{X}$, and its ratio \lxlbol\ to the bolometric luminosity). 
Two nonparametric rank-correlation tests (Spearman's~$r_\mathrm{S}$ and Kendall's~$\tau$;
\citealt{nr77}) were used. These robust tests allow us to calculate
correlation coefficients and to obtain a two-sided (correlation or
anticorrelation) significance for the \emph{absence} of correlation. Thus, a low
correlation coefficient ($r_\mathrm{S}$ or $\tau$) can be associated with a high probability
that the sample is not correlated. Note that Kendall's~$\tau$ is
more nonparametric than Spearman's $r_\mathrm{S}$ because it uses only the relative
ordering of ranks instead of the numerical difference between ranks
\citep{nr77}. No uncertainty was included in the tests. In
Table~\ref{tabcorr}, we give the rank-correlation coefficients
together with their two-sided significances in parentheses.
We have defined two data groups for the tests. The first group (hereafter DG1) 
corresponds to power-law indices or flare rates derived from ML fits to the
cumulative distributions, while data group 2 (hereafter DG2) corresponds to 
indices or flare rates derived from fits to the differential distributions.

\subsection{Correlations of the Power-Law Index $\alpha$}
\subsubsection{Coronal Luminosity $L_\mathrm{X}$}\label{correlalpha}
Coronal luminosities (see eq.~[\ref{lx}]) were used to test their 
correlation with $\alpha$ (Fig.~\ref{lxcor}). For each data group, the 
significance levels (Table~\ref{tabcorr}) are usually smaller than 5~\%, the highest level
reaching 10~\%. Therefore, placing a limit of 5~\% to the two-sided significance 
level, the correlation between $\alpha$ and $L_\mathrm{X}$ is marginally
significant at best. Such a correlation can be explained as
follows. For saturated stars ($L_\mathrm{X} \approx 10^{-3} L_\mathrm{bol}$), the X-ray
luminosity should decline for stars with spectral type from F to M (hence increasing
color index \bv) because of the decrease of their bolometric luminosity. In
section~\ref{sptyp}, we have found a suggestion that the power-law index $\alpha$ is
weakly correlated with the stellar spectral type. Therefore, a weak correlation of $L_\mathrm{X}$
with $\alpha$ can be expected. Note that due to the scatter in the \lxlbol\ ratio in our
sample, this can lead to a scatter in the dependence of $L_\mathrm{X}$ on the spectral
type, hence in the dependence of $L_\mathrm{X}$ on the index $\alpha$.

\placefigure{lxcor}

\subsubsection{Ratio \lxlbol}
We have tested the correlation between \lxlbol\ and the power-law index. The bolometric
luminosities were calculated\footnote{$\log L_\mathrm{bol} = \log L_\mathrm{bol,\odot} + 0.4 (
M_\mathrm{bol,\odot} - M_\mathrm{bol})$, where $M_\mathrm{bol} = V - ( 5 \log d - 5 ) + BC$. Here, $M_\mathrm{bol,\odot} =
4.64$~mag, $L_\mathrm{bol,\odot} = 3.85 \times 10^{33}$~\ergps\ and $BC$ is taken
from \citet{schd82}.} from parameters in Table~\ref{list} and corresponding
bolometric corrections. Two-sided significance levels
(Table~\ref{tabcorr}) show that the correlation between \lxlbol\ and $\alpha$ is
not significant.

\subsubsection{Rotation Period $P$}
For stars with known rotation periods (Table~\ref{list}), we have tested the
correlation between the power-law index $\alpha$ and $P$. Note that for
$\kappa$~Cet, for which there are two data sets, we have used weighted
means\footnote{$\alpha = \left( w_1 \alpha_1 + w_2 \alpha_2 \right) / \left( w_1
+ w_2 \right)$, where $w_i = 1/\left(\sigma_{i,\mathrm{u}}\sigma_{i,\mathrm{l}}\right)$, and
$\sigma_{i,\mathrm{u}}$ and $\sigma_{i,\mathrm{l}}$ are upper and lower error bars, respectively. For
power-law indices of DG1, we have $\sigma_{i,\mathrm{u}} = \sigma_{i,\mathrm{l}} = \sigma_{i}$.} 
of the indices $\alpha$. Our final sample then comprised 7 data points.  From
the two-sided significance levels, we can state that no
correlation between the power-law index $\alpha$ and the rotation period $P$ is
present in our sample (Table~\ref{tabcorr}).

\subsubsection{Projected Rotational Velocity $v \sin i$}
Projected rotational velocities from Table~\ref{list} were used. We used the same
procedure as above to calculate the weighted mean of $\alpha$ for $\kappa$~Cet; 
for the tests, the upper limits (GJ~411, CN Leo) have been omitted. Our sample
then contained 8 data points for each data
group. The two-sided significances for the present correlation tests
(Table~\ref{tabcorr}) imply the absence of a significant correlation between the
power-law index $\alpha$ and the projected rotational velocity $v \sin i$ in our
data sample.

\subsubsection{Rossby Number $R_0$}
We have used the available periods in Table~\ref{list} and have calculated the
convective turnover times $\tau_\mathrm{c}^{\left(2\right)}$ (the number 2 refers to the ratio of
the mixing length to the scale height) from the \bv\ color index and equation~(4) of
\citet{noy84} in order to derive $R_0$. The nonparametric tests again 
suggest an insignificant correlation between the index $\alpha$
and the Rossby number.

\placetable{tabcorr}

\subsection{Correlations of the Flare Occurrence Rate}
\subsubsection{Flare Rate vs. $L_\mathrm{X}$}
Figure~\ref{distrlx} shows the occurrence rate of flares with energies larger 
than $10^{32}$~ergs ($E_\mathrm{c}$) versus the coronal luminosity $L_\mathrm{X}$. 
Quiescent X-ray luminosities (corrected to an energy range between 0.01 and 
10 keV) taken from the literature \citep{coll88,pall90,hun98,hun99} were also 
used for comparison (crosses), although they were not taken into account in the following 
tests. The evident correlation between $N(>E_\mathrm{c})$ and $L_\mathrm{X}$ is confirmed by
Spearman's $r_\mathrm{S}$ and Kendall's~$\tau$ tests. The former has rank-correlation coefficients of $0.95$ and $0.94$, 
with two-sided significances for deviation from zero of $2 \times 10^{-6}$ and $4\times 
10^{-6}$ for DG1 and DG2, respectively. The latter test has coefficients of $0.85$ and 
$0.82$, while significances are $1 \times 10^{-4}$ and $2 \times 10^{-4}$. 
This implies that a correlation between $L_\mathrm{X}$ and the flare occurrence rate 
is highly significant for our data sample (Figure~\ref{distrlx}).  
The linear best-fit in the $\log - \log$ plane for DG1 is 
$\log N(>E_\mathrm{c}) = (-26.7 \pm 2.9) + (0.95 \pm 0.10)
\log L_\mathrm{X}$ (number of flares per 
day), while the best-fit for DG2 was $\log
N(>E_\mathrm{c}) = (-25.5 \pm 2.8) + (0.90 \pm
0.10)\log L_\mathrm{X}$, hence the relation between the flare rate and the luminosity is
compatible with proportionality. Note that, as the correlation between $\alpha$ and $L_\mathrm{X}$
was marginal at best (section~\ref{correlalpha}), we can safely state that proportionality
exists between $L_\mathrm{X}$ and the normalization factor $k_2$ (and hence $k_1$).

\placefigure{distrlx}

\subsubsection{Normalized Flare Rate vs. \lxlbol}\label{flracc}
The canonical saturation limit for 
stars with different spectral types has been found to appear at different $v \sin i$
\citep[e.g.,][]{caill85,stauff94,rand96,stauff97}, and therefore, based on the relation of 
\citet{pall81} between $v \sin i$ and $L_\mathrm{X}$ for unsaturated stars, at different X-ray 
luminosities. Figure~\ref{lxlb} shows the coronal luminosity $L_\mathrm{X}$ against 
the bolometric luminosity $L_\mathrm{bol}$. It emphasizes the different loci of our coronal sources 
with respect to saturation. Note that \lxlbol\ ratios range from $\approx$$10^{-5}$ to 
$\approx$$10^{-3}$. For our correlation tests, we have normalized, for each source, the occurrence rates of flares with
energies larger than $10^{32}$~ergs with the occurrence rate $N(>E_\mathrm{c})_\mathrm{sat}$ \emph{at the saturation
turn-on for the given spectral type} ($L_\mathrm{X,sat}=10^{-3}L_\mathrm{bol}$) derived from the best-fits
to $N(>E_\mathrm{c})$~vs.~$L_\mathrm{X}$ in the previous section. Thus, we are able
to check whether or not the normalized 
flare occurrence rate stays constant at unity at activity saturation. Figure~\ref{distrlxlb} suggests
that it does and that the flare rate saturates at activity saturation. In the
following sections, we will suggest that this effect is not biased by the 
absence of stars ``beyond'' saturation. Note a few discrepant features, such as for 47 Cas (point 2). Its point in DG2
is about 1/2 dex higher than in DG1. Since its index is larger ($\alpha \approx 2.6$)
for DG2 than for DG1 ($\alpha \approx 2.2$), and since its minimum observed
flare energy is about $10^{33}$~ergs, it follows that
$N(>E_\mathrm{c})$ is larger for DG2 than for DG1.
Similarly, CN Leo (points 11~\&~12) has two different $\alpha$ indices for the 1994 and 1995 observations ($\alpha
\approx 2.2\textrm{ and }1.5$). This large discrepancy (due to the flat low-energy end of
the 1995 distributions) induces different normalized flare occurrence rates.

\placefigure{lxlb}
\placefigure{distrlxlb}

\subsubsection{Normalized Flare Rate vs. Normalized $v \sin i$}
As before, it was necessary to normalize the flare
occurrence rates. In section~\ref{flracc}, we
have shown that, for our sample, the normalized flare occurrence rate
does not show a trend to increase at saturation. However, our sample contains stars ``beyond''
saturation, i.e., stars that appear saturated ($L_\mathrm{X}/L_\mathrm{bol} \approx
10^{-3}$) but that rotate faster than a star at the onset of saturation.
Therefore, we have normalized $v \sin i$ with $(v \sin i)_\mathrm{sat}$, where the latter
were obtained from the \citet{pall81} relation ($L_\mathrm{X} = 1.4 \times 10^{27} [v
\sin i]^{1.9}$) at the saturation turn-on for each stellar spectral type. Stars
at the saturation level thus have normalized velocities around 1, while those
beyond that level have values significantly higher than 1
(Fig.~\ref{distrvsini}). In our sample, two stars show high values. The first
is the bright K1 dwarf AB Dor, and the second is the M6 dwarf CN Leo,
although its projected rotational velocity is an upper limit. Hence, AB Dor is
the only star that supports the suggestion of constant normalized flare rates at
$v \sin i$ saturation. However, in the following section, we will show that the
result is also supported by a correlation with the Rossby number $R_0$, which is
less dependent on spectral type. We have performed
a fit to the data, assuming a saturation function of the type 
\begin{equation}
\frac{N(>E_\mathrm{c})}{N(>E_\mathrm{c})_\mathrm{sat}} = 1 - \exp \left( - \frac{1}{\zeta} \frac{v \sin i}{(v \sin i)_\mathrm{sat}}\right),
\label{sateq}
\end{equation}
with $\zeta$ as the single fit parameter. We derived $\zeta = 0.56$ for DG1, and
$\zeta = 0.33$ for DG2. For the fits, we have used the estimated equatorial velocity for
47 Cas, the logarithmically averaged flare rate for $\kappa$~Cet, and we have
not taken into account the upper limits for CN Leo and GJ 411.

\placefigure{distrvsini}

\subsubsection{Normalized Flare Rate vs. $R_0$}
We have further tested the saturation of the flare rate using $R_0$. 
Compared to the normalized projected
rotational velocity, the Rossby number does not contain the uncertainty due
to the projection angle $i$. Furthermore, it does not need to be normalized
as it already corresponds to a normalized rotation period. However, the rotation
period is not available for each star of our sample. Figure~\ref{distrross} (upper and
middle panels) suggests a saturation effect, although again only AB Dor
supports it. Also, the $N(>E_\mathrm{c})/N(>E_\mathrm{c})_\mathrm{sat}$ vs. $R_0$ plots 
(upper and middle panels) are very similar to the well-known
activity saturation relation with the Rossby number (lower panel of
Figure~\ref{distrross}). Similarly to the normalized flare rate saturation, the
luminosity saturation appears at $\log R_0 < 1.2$. Both the normalized flare rate and the luminosity
decrease above this limit, as previously found for \lxlbol\ \citep[e.g,][]{rand96,stauff97}. 
Two lines were overlaid for $\log R_0
\ge 1.2$. The solid line corresponds to the best fit to the data of 
\citet[$\log L_\mathrm{X}/L_\mathrm{bol} = -4.4 - 1.12 \log R_0$]{rand96}, while the
dashed line corresponds to our best-fit solution, 
$\log L_\mathrm{X}/L_\mathrm{bol} = (-4.76 \pm 0.30) - (1.72 \pm 0.42)
\log R_0$, with the ratios \lxlbol\ of $\kappa$~Cet being logarithmically
averaged. Our sample follows approximately the relation of
\citet{rand96}. Hence, together with the upper and middle panels, the lower panel
of Figure~\ref{distrross} reinforces the suggestion that there is a
saturation of the flare rate at the activity saturation.

\placefigure{distrross}

\subsection{Flare Power vs. $L_\mathrm{X}$}
In Figure~\ref{en}, we have plotted the X-radiated power $P_F$ from the detected
flares as a function of the average luminosity $L_\mathrm{X}$. There is an
obvious correlation (Spearman: $r_\mathrm{S} = 0.93$, $P[r_\mathrm{S} = 0]=1.2 \times 10^{-5}$;
Kendall: $\tau = 0.82$, $P[\tau = 0] = 2.1 \times 10^{-4}$) between the parameters. This
again demonstrates the importance of the flare contributions to the
observed radiation from active coronae. Note that, for our sample and
for our flare detection threshold, \textit{about 10~\% of the X-ray luminosity originates
from detected flares}.
 
\placefigure{en}

\section{DISCUSSION AND CONCLUSIONS}\label{sectdisc}

In the present work we have investigated statistical properties of EUV flare events
on time scales of days and weeks. Although the sensitivity of the \euve\ DS instrument is
quite limited, it provides long time series  that are sufficient to
draw rough conclusions on the statistical flare behavior of active stellar coronae.  

The first aspect of interest is the distribution of flare energies. In the solar
case, X-ray flares are distributed in energy according to a power law with
a power-law index around 2 \citep[e.g.,][]{crosby93}. Many of the active stars studied
here are
quite different from the present-day Sun, with distinct coronal behavior. For example, 
high-energy particles are continuously present in their coronae as inferred from their
steady gyrosynchrotron emission \citep[e.g.,][]{linsky83,guedel94}; quiescent coronal temperatures reach values of
20~MK which are typical on the Sun only in rather strong flares. Individual
flare energies accessible and observed by \euve\ in this study are, in several cases, 
not observed on the Sun at all. Since a limited amount of magnetic energy is available in
the reconnection 
zones of the coronal magnetic fields, one would even expect some upper threshold to 
observable flare energies from solar-like stars (and therefore  to the observed flare 
X-ray radiation).

Our investigation on active main-sequence stars has shown that, for the energy
range observed, the flare occurrence rate 
distributions in energy can be fitted by power laws. We have not found significant
evidence for
broken power laws that indicate a threshold energy. The largest amount of radiated energy
was found to be $10^{35}$~ergs in our flare sample, exceeding the X-ray output of very
large
solar flares by two to three orders of magnitude.   

Measuring the value of the power-law indices of the distributions is pivotal for assessing
the role
of flares in coronal heating. The quite limited statistics make  conclusions 
somewhat tentative, although we emphasize the following.
The power-law indices  definitely  cluster around a value of 2; they may be slightly
different for different stellar spectral types (Table~\ref{tabresult}). The stellar
distributions
are thus broadly equivalent to solar distributions, which implies that a) the cause of
flare initiation in
magnetically very active  stars may be similar to the Sun, and
 b) that the trend continues up to energies at least two orders of magnitude  higher
than observed on the Sun. We are conversely motivated to  extrapolate  our distributions
to
lower energies  given the rather large range over which solar flares follow a power law. 
Caution is in order, however, toward low-energy flares for which the distributions may
steepen \citep{krucker98}.  

\citet{hudson91} argued that, for a power-law index above 2, an extrapolation
to small flare energies could explain
the radiated power of the solar  corona. We derived
(eq.~[\ref{emin}]) the minimum flare energies $E_\mathrm{min}$ required for the power laws
to explain our stellar X-ray luminosities (Table~\ref{tabresult}). 
For stars with $\alpha >2$ or just barely below $2$, minimum flare energies  
around $10^{29}-10^{31}$~ergs were  obtained. Such energies correspond to intermediate
solar flares. Explicit measurements of flare energies below our detection thresholds will however
be required  to conclusively estimate their contribution to the overall radiation.

We have found a trend for
a flattening of the flare rate distributions in energy toward later spectral types. 
F and G-type stars tend toward power-law indices $>2$, while K and M dwarfs
tend toward indices $<2$.
If supported by further, more sensitive surveys, it suggests that flares play a more
dominant role in the heating of F and G-type coronae, while they  cannot provide
sufficient energy to explain the observed radiation losses in K and M dwarfs. On the other
hand,
part of this trend could be due to the bias introduced by the identification method and the 
length of the GTIs. Although not found in our data, later-type (K and M-type) stars
may show flares 
that are typically shorter than those of G dwarfs, partly because of the smaller
distances of the stars that give access  to less energetic (and therefore, as on the Sun,
typically
shorter) flares. But then  the flare duration is smaller than the typical GTI gaps
(about 3000~s) so that flares that occur between
the GTIs remain completely undetected. This effect can considerably
flatten the flare rate distributions.

Given that our study is restricted to  flares with energies typically exceeding
$10^{31}-10^{32}$~ergs, it is little suprising that the observed flare radiation 
amounts to only a fraction of the total EUV and X-ray losses.
We infer an (observed) fraction of approximately 10~\% relative to the average
(quiescent) coronal luminosity.  This lower limit will undoubtedly increase
with better instrument sensitivity. Our study therefore clearly indicates that flares 
provide an important and significant contribution to the overall heating of active
stellar coronae.

We have further explored whether statistical flare properties are correlated with some
physical properties of the stars, such as activity indicators and rotation
parameters. The power-law index $\alpha$ does not correlate
with any of the rotation parameters ($P$, $v \sin i$, $R_0$) nor with the ratio
\lxlbol. The absence of clear correlations suggest that the activity phenomena
related to flares are similar on stars of all activity levels. A marginally significant correlation with the coronal luminosity
was found. This result is probably related to the trend found for the dependence 
of $\alpha$ on the stellar spectral type. 

On the other hand, the flare occurrence rate above a given lower energy threshold
is correlated with each of the activity indicators and rotation parameters. A single power
law
\mbox{($N[>E_\mathrm{c}] \propto L_\mathrm{X}$)} fits the correlation between flare rate and  the
coronal luminosity quite well, indicating that energetic flares occur more frequently in X-ray 
luminous stars than in X-ray weak stars. In order to compare the activity levels
between stars of our sample, we have normalized the flare rate to its value that a
star adopts at its saturation level. The normalized flare occurrence rate increases with
increasing activity but stays constant for saturated stars. Flare rate saturation
underlines the close relation between flares and the overall ``quiescent'' coronal emission.
 We now  ask more specifically whether
the apparently  quiescent X-ray radiation could be related to the derived flare
distributions.

In simple terms, we expect a larger magnetic filling factor on  magnetically more active
stars, or more numerous (or larger) active regions than on low-activity stars.
A higher filling factor naturally implies a proportionally higher quiescent X-ray 
luminosity and proportionally more numerous flares, so that we expect a linear
correlation between $L_\mathrm{X}$ and the flare rate. But what is the nature of the
quiescent emission?

It is known from X-ray observations that
the average quiescent coronal temperatures characteristically increase with increasing
activity. The concept of an average coronal temperature is, however, 
somewhat problematic. \citet*{schrijver84} used single-$T$ fits to
\textit{Einstein}/IPC data
that roughly imply that the total stellar volume emission measure in X-rays
$\mathrm{EM_\star}$ is proportional to $T^3$ (as discussed in \citealt*{jordan91}).  
Since the radiative cooling function $\Lambda(T)$ in the range of interest ($3-30$~MK)
scales approximately  like $T^{-\phi}$ with $\phi \approx 0.3$ \citep*{kaastra96}, 
we have for $L_\mathrm{X}$
\begin{equation}
L_{\rm X} \approx {\rm EM_\star} ~ \Lambda(T) \propto {\rm EM_\star} ~ T^{-\phi}
\label{lxtrel1}
\end{equation}
and hence $L_\mathrm{X} \propto T^{2.7}$. Considering that the typical coronal emission
measure is distributed in temperature, multi-$T$ fits
appear to better represent average coronal temperatures.  
Also, to disentangle functional dependencies from other
stellar parameters (e.g., the stellar radius), a uniform sample
of stars should be used.
\citet{guedel97} derived two-temperature models from \textit{ROSAT} data for a sample of stars 
that  differ only in their activity levels but are otherwise analogs to the Sun. 
For the hotter component, they
found $L_\mathrm{X} \propto T_{\rm hot}^{4.16-5.09}$, the range of the exponent illustrating
two different spectral models applied. If, however, we compute the mean of the
two temperatures weighted with the corresponding emission measures from their Table 3,
we find $L_\mathrm{X} \propto T^{4.8}$ independent of the spectral model. 
The  two more active stars for which \textit{ASCA} 3-$T$ fits were available in \citet{guedel97}
agree well with this trend (for cooler coronae, \textit{ASCA} is not sufficiently sensitive
to derive a coronal emission measure distribution). Note that the 3-$T$ fits
in turn represent the derived emission measure distributions in \citet{guedel97}
quite well. We therefore conclude that the emission-measure weighted average
coronal temperature roughly scales as $L_{\rm X} \propto T^{4.8}$, although with a 
considerable uncertainty in the exponent.

\citet{hearn75} and 
\citet{jordan87} discuss ``minimum flux'' coronae that follow  a relation 
$\mathrm{EM_\star} \propto T^3 g_\star$ where $g_\star$ is the stellar surface gravity. 
In a similar way,  \citet*{rtv} find scaling laws for 
closed static magnetic loops that relate  external heating, loop pressure, loop-top temperature, and
loop length. The two scaling laws combine to 
\begin{equation}
\dot{E}_{\rm H} \propto {T^{21/6}\over L^2} \label{rtvheating}
\end{equation}
where $\dot{E}_{\rm H}$ is the heating rate, $T$ is the (dominating) loop-top temperature,
and $L$ is the loop length. \citet{schrijver84}, based  on these scaling laws, conclude
that different families of loops must be present on active stars. 
In either case, explaining the locus of the measured ($T$, $\mathrm{EM_\star}$) on the empirical relation
requires an explanation for a specific amount of heating energy input to the system.
Our investigation suggests that flare events contribute significantly to the
observed overall emission. If the concept of a truly quiescent radiation is
abandoned completely for magnetically active stars, then we may ask whether flares
themselves can explain the observed relation between $\mathrm{EM_\star}$ and $T$ while at the same time 
accounting for the proportionality between flare rate and $L_\mathrm{X}$. We briefly discuss
two extreme cases:

i) {\it Filling factor-related activity.} We first assume that flares are statistically
independent and occur at many different, unrelated flare sites. The quiescent emission
in this case is  the superposition of all flare light curves. The proportionality
between flare rate and $L_\mathrm{X}$ is simply due to more numerous flare loops
on stars with higher magnetic filling factors (and hence higher activity), in which
the explosive energy releases build up the observed loop emission measure.
We ask whether the statistical flare distribution determines 
the average coronal temperature as well. 

From our flare samples, we have found that the flare duration does not obviously depend on
the flare energy (as typically found for impulsive flares on the Sun; \citealt{crosby93}). 
We therefore first assume one fixed time constant for all flares. 
For example, radiatively decaying flares 
with a similar evolution of coronal densities, abundances and temperatures show similar 
decay time scales. In this case, the  {\it single flare peak} luminosity 
$\mathcal{L}_{\rm X}$ is proportional to the total radiated flare energy $E$. 
Further, for solar (and stellar) flares, a rough relation between \emph{peak} emission measure 
$\mathrm{EM}$ and \emph{peak} temperature $T$ has been reported by \citet*{feldman95}.
For the interesting range between $5-30$~MK, the relation can be approximated by
\begin{equation}
\mathrm{EM}  = aT^b \quad {\rm [cm^{-3}]} \label{emtrel}
\end{equation}
with $a \approx 10^{12}$~cm$^{-3}$, $b \approx 5\pm 1$, and  $T$ measured in K. 
We approximate the flare contribution to the radiative losses by the values around flare
peak. For the flare radiative losses, we have, equivalent to equation~(\ref{lxtrel1}),
\begin{equation}
\mathcal{L}_{\rm X} \approx {\rm EM} ~ \Lambda(T) \propto {\rm EM} ~ T^{-\phi}
\label{lxtrel2}
\end{equation}
again with $\phi \approx 0.3$ over the temperature range of interest. 
With equations~(\ref{emtrel}) and (\ref{lxtrel2}),
we obtain 
\begin{equation}
E \propto \mathcal{L}_{\rm X} \propto T^{b-\phi}. \label{etrel}
\end{equation}
To derive a characteristic, emission-measure weighted, time-averaged mean coronal temperature $\overline{T}$,
we average over all flare temperatures $\left[ T_\mathrm{s}, T_0 \right]$ by using their
peak $\mathrm{EM}$ and their occurrence rates as weights:
\begin{equation}
\overline{T} =  \frac{\int_{T_\mathrm{s}}^{T_0}T ~ {\rm EM(T)} ~ dN/dE ~ dE/dT ~ dT}
              {\int_{T_\mathrm{s}}^{T_0}  {\rm EM(T)}~ dN/dE ~ dE/dT ~
dT}.\label{averageT}
\end{equation}
Here, $T_\mathrm{s}$ represents the typical temperature of the smallest contributing
flares. From equations (\ref{emtrel})--(\ref{averageT}), we obtain 
\begin{equation}
\overline{T} = \frac{z}{z+1} ~ T_0 ~ \frac{1 - (T_\mathrm{s}/T_0)^{z+1}}%
                {1 - (T_\mathrm{s}/T_0)^{z}}
                \label{averageT2}
\end{equation}
where 
\begin{equation}
z = (2-\alpha)b - (1-\alpha)\phi.
\end{equation}
Without loss of generality, we assume  $T_\mathrm{s} = 1$~MK. The upper temperature 
limit $T_0$ corresponds to 
the largest flare energy $E_0$ typically contributing to the apparently quiescent
emission. Since the flare rate is given by the power law (eq.~[\ref{k1def}]), the
characteristic
value for $E_0$ scales like $k_1^{1/\alpha}$, i.e., with equation (\ref{etrel}), 
$T_0 \propto k_1^{1/(\alpha(b-\phi))}$. The normalization factor $k_1$ is proportional to
the  overall stellar X-ray luminosity $L_\mathrm{X}$ for given $\alpha$.
The steepest dependence of $\overline{T}$
on $k_1$ is obtained in the limit of small $\alpha$, i.e. $z > 0$, so that $\alpha <
(2b-\phi)/(b-\phi)$,
implying $\alpha \la 2.1$. Then, for $T_0 \gg T_\mathrm{s}$, we find
$\overline{T} \propto T_0$, i.e., $L_{\rm X} \propto \overline{T}^\beta$ with
$\beta = \alpha(b-\phi)$. For reasonable values of $\alpha = 1.5 - 2.1$, we find
$\beta \approx 5.5-12$, i.e., a dependence that is at least somewhat steeper than observed.

We can repeat the derivation under the assumption of some dependence between flare energy
$E$ and duration $D$. Typically, larger flares last longer. The function
$D(T) = D(E[T])$ should then be used as an additional weighting factor in 
equation~(\ref{averageT}). Assuming that $D 
\propto E^{1/2}$ (e.g., in the case of an
energy-independent ``flare curve shape''), we find $E \propto \mathcal{L}_\mathrm{X}^2$. In that
case, equation~(\ref{averageT2}) remains valid, while 
$z = (4-2\alpha)b-(3-2\alpha)\phi$ and $L_\mathrm{X} \propto \overline{T}^\beta$
with $\beta = 2\alpha(b-\phi)$ for ``small'' $\alpha$, hence a steeper dependence on 
$\overline{T}$ than for the first case.

ii) {\it Loop reheating.} In the other extreme case, the flares repeatedly reheat the
same coronal plasma in certain ``active'' loop systems \citep[e.g.,][]{kopp93}. In that case,  for 
most of the time the coronal loops fulfill a quasi-static approximation  equivalent
to the loop scaling law given by \citet{rtv} \citep[see][]{jakimiec92}. The situation is 
equivalent to steadily heated loops.  The first scaling law of \citet{rtv}, 
$T_{\rm max} \propto (pL)^{1/3}$, implies EM $\propto T^4$
and hence $L_{\rm X} \propto T^{3.7}$ for a given loop length with $T$ in the range of
interest. For loops exceeding the coronal pressure scale height, \citet{schrijver84}
give an expression EM $\propto T^5$ (with other parameters fixed), and hence
$L_{\rm X} \propto T^{4.7}$. In this limit, we attribute the higher temperatures 
of more active stars to higher reheating rates in a similar number of loops rather than to a
larger number of statistically independent heating events in more numerous active regions 
or loops. The cause of an enhanced reheating rate on more active stars in a similar number 
of loop systems remains to be explained, however.

Comparing the observed average coronal temperatures with the two extreme results,
we see that the observed trend lies in the middle between the extreme values. 
Flare heating is thus a viable candidate to explain the trends seen in $L_\mathrm{X}$ and 
in $T$, although it is not conclusive whether flares occur independently or
whether they act as reheating agents of coronal loops. It appears unlikely that either extreme
is appropriate. While a larger magnetic filling factor on more active stars
undoubtedly produces more numerous active regions and thus a higher flare rate,
the higher coronal filling factor is also likely to lead to more numerous
reheating events. This is compatible with the observed trend between
$L_{\rm X}$ and $\overline{T}$. 

This study has shown that flares  can provide a significant amount of energy to
heat the coronae of active stars. Although the definite answer to which
mechanism is responsible for coronal heating is not yet available, our data
sample suggests that flares are promising contributors. 
Better sensitivity together with uninterrupted observations (such as those  
provided by the new generation of X-ray satellites \textit{XMM-Newton} and
\textit{Chandra}) will be needed to solve part of the mystery. Our method is
limited to flares explicitly detectable in the light curves. Alternative methods
are available that model light curves based on statistical models. These
investigations are the subject of a forthcoming paper \citep{drake00}.

\acknowledgements
We are grateful to Keith Arnaud for his help and for providing his software that
allowed us to perform simultaneous fits to the differential distributions. We
also thank Damian Christian for his help in the \euve\ data reduction. M.~A. acknowledges
support from the Swiss National Science Foundation, grant 21-49343.96, from
the Swiss Academy of Sciences and from the Swiss Commission for Astronomy.

\clearpage

\figcaption[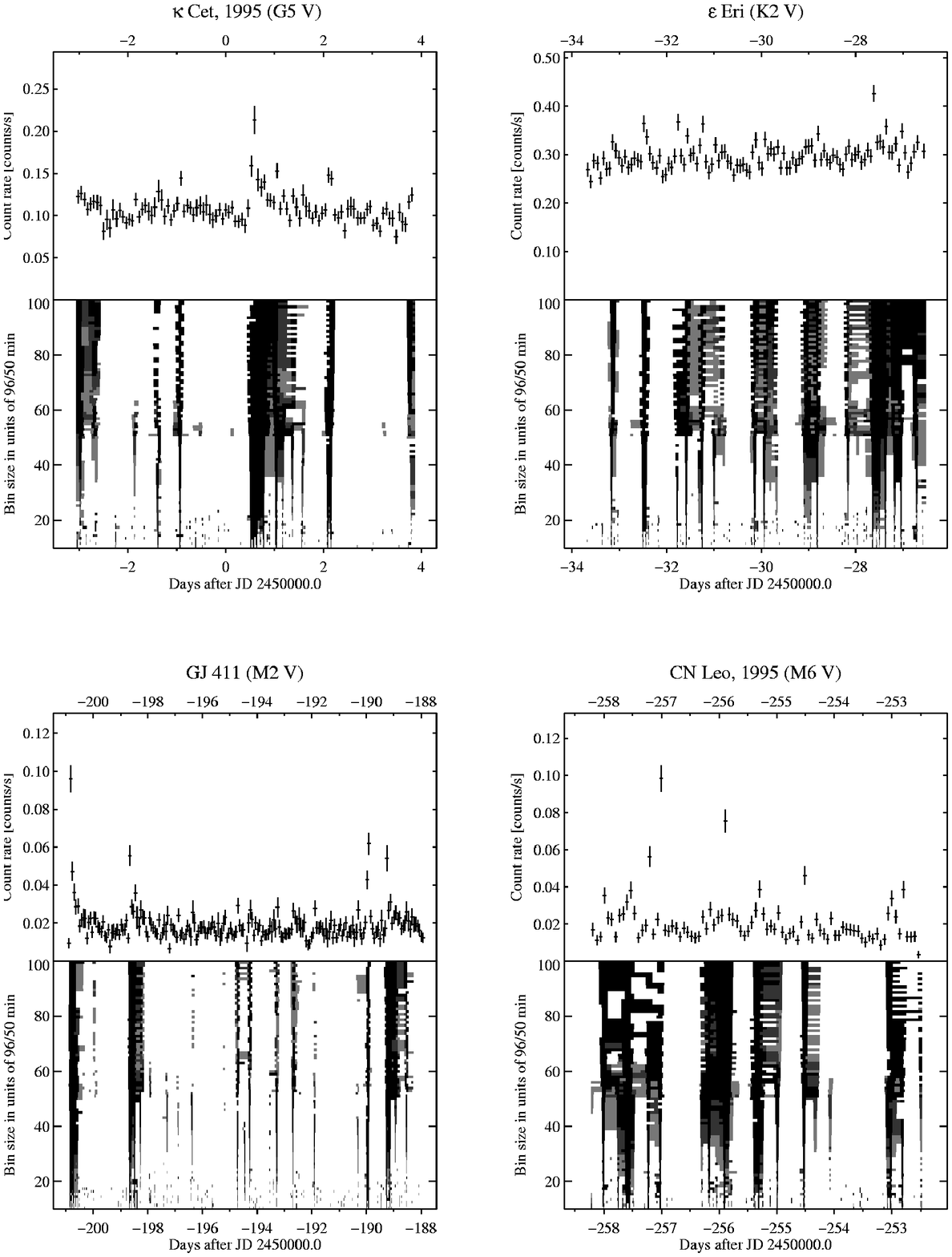]{Examples of light curves and significance plots.\label{lcsigdef}}
\figcaption[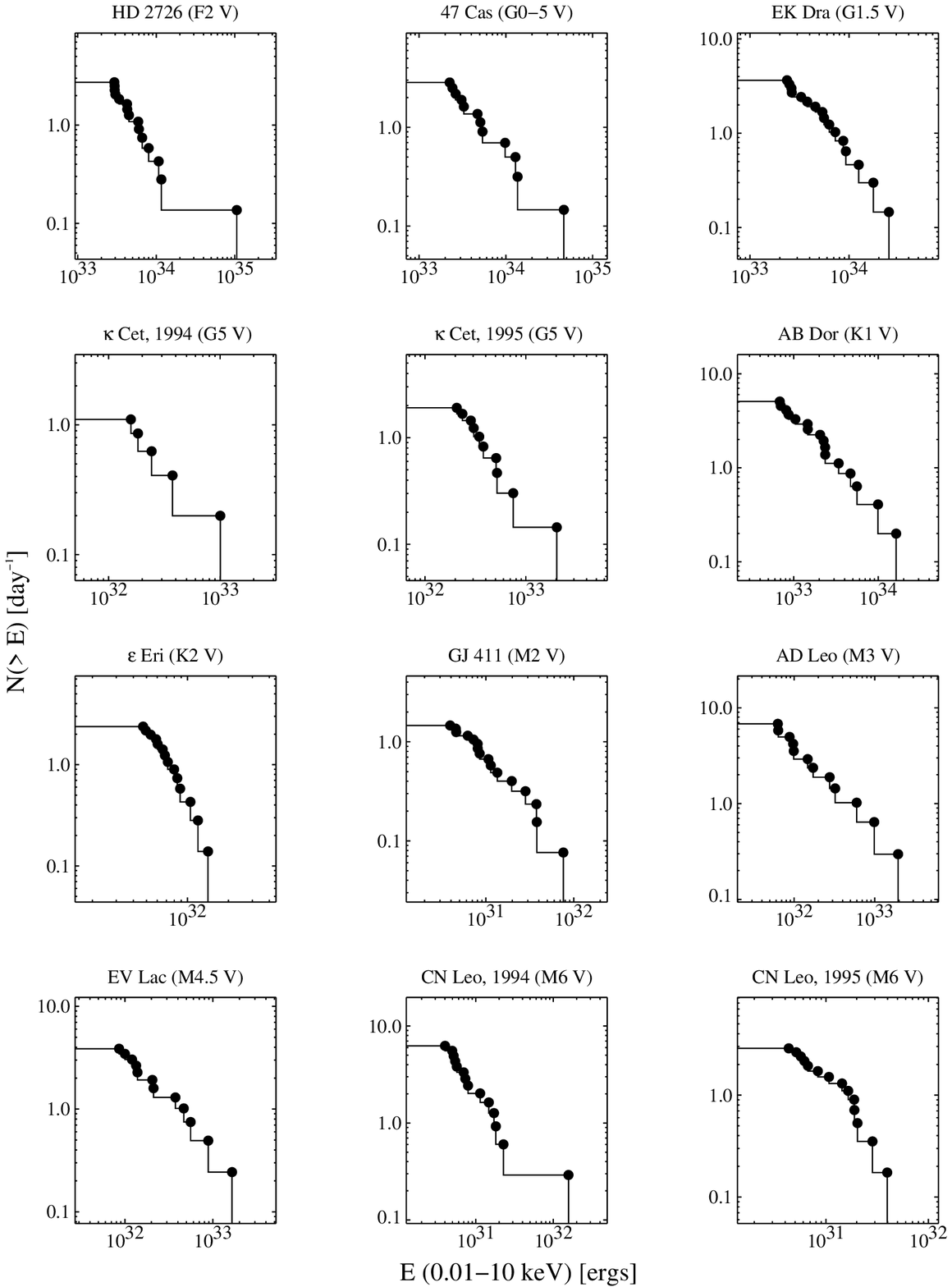]{The cumulative flare occurrence rate distributions for all twelve
data sets.\label{cumdis}}
\figcaption[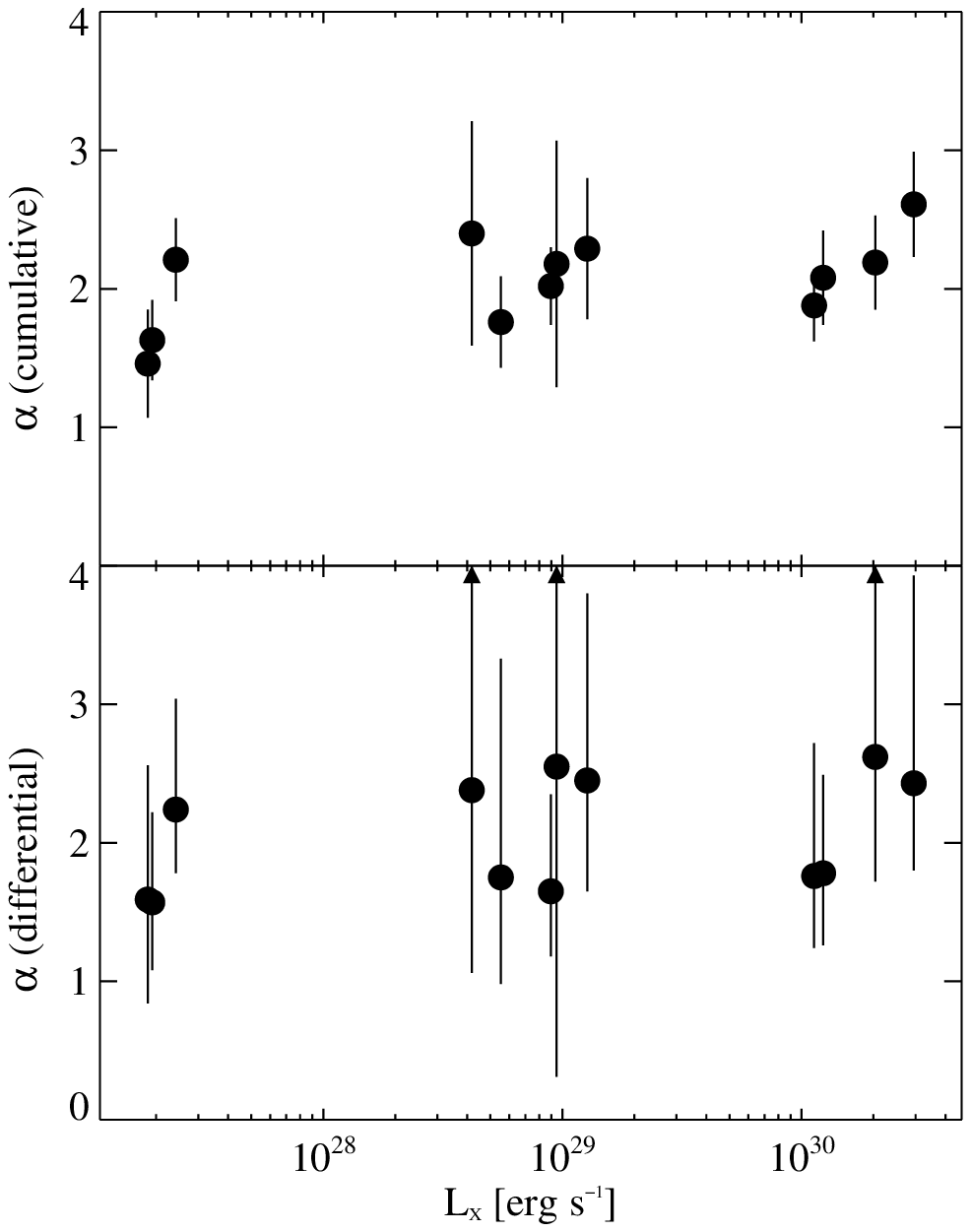]{Power-law index $\alpha$ vs. coronal $L_\mathrm{X}$.\label{lxcor}}
\figcaption[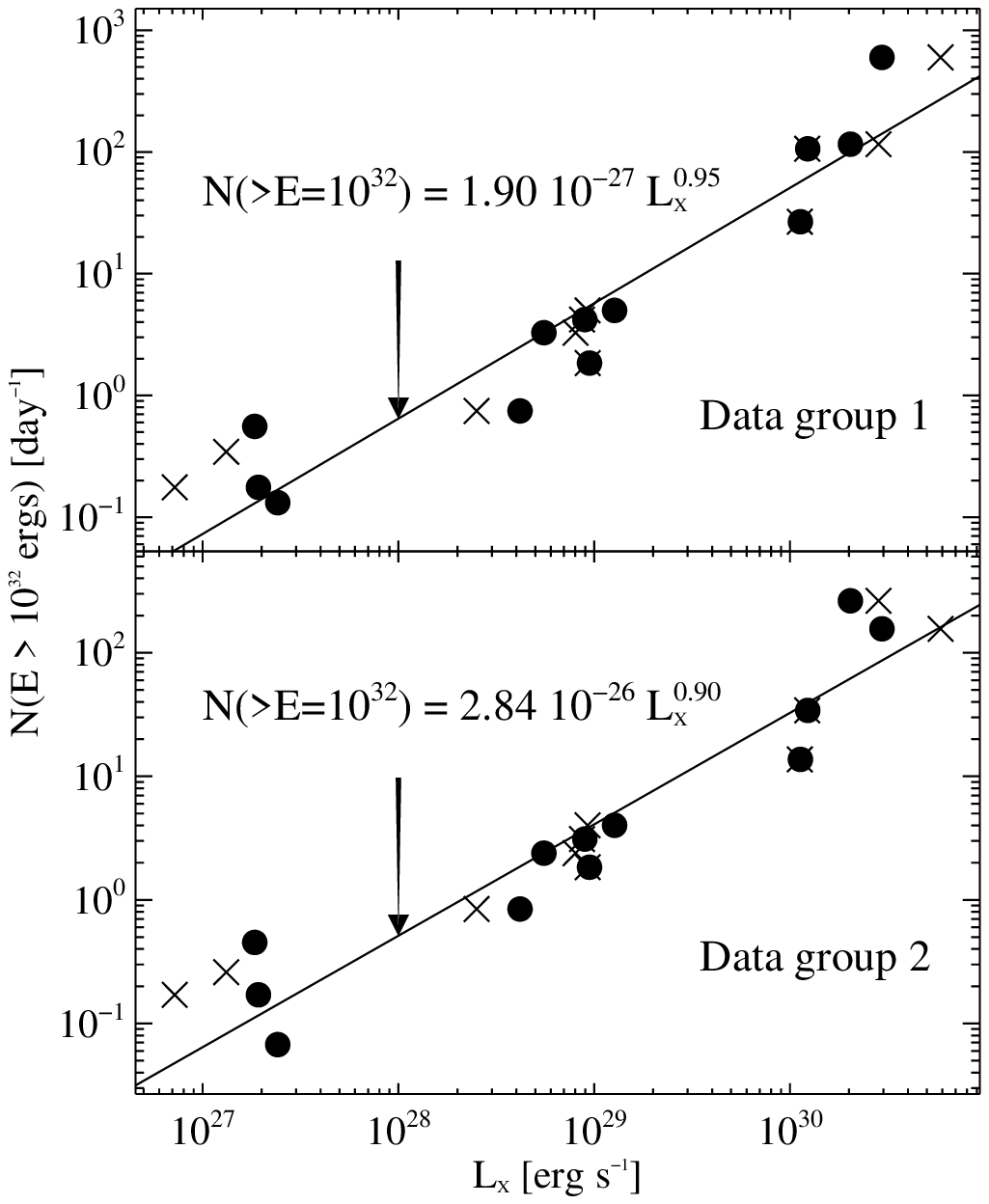]{Occurrence rate of flares with energies larger than $10^{32}$~ergs
vs. coronal luminosity $L_\mathrm{X}$. Data group 1 corresponds to the rates 
derived from the fits to cumulative distributions, while data group 2 corresponds to 
the rates derived from the fits to differential distributions. The $\chi^2$ linear best-fits
in the $\log - \log$ plane are shown as straight lines together with the analytical 
formulation. Crosses represent $L_\mathrm{X}$ values for these sources taken from the 
literature (see text).\label{distrlx}}
\figcaption[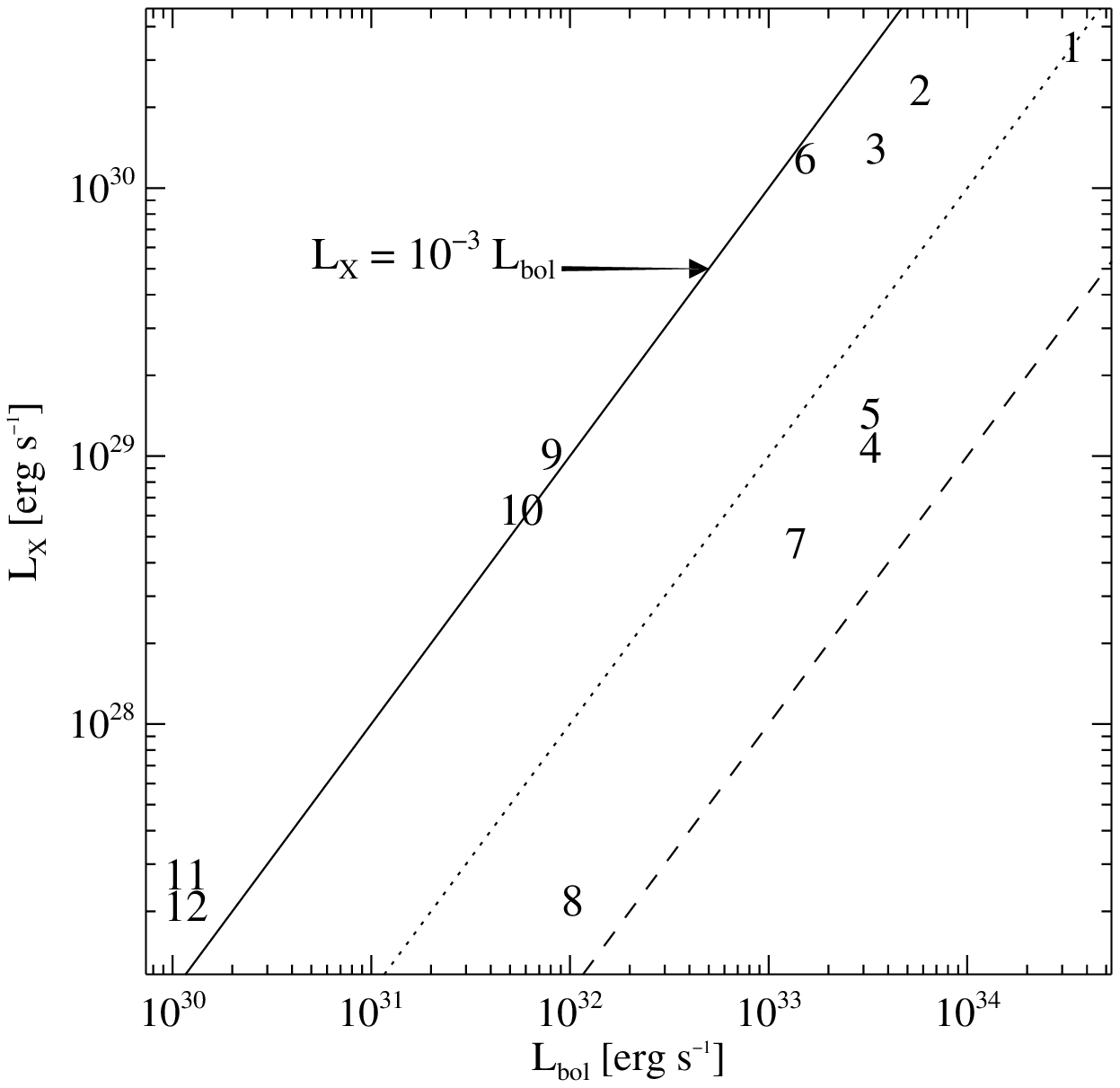]{Coronal luminosity $L_\mathrm{X}$ vs. bolometric luminosity
$L_\mathrm{bol}$. The source identification numbers refer to Table~\ref{list}. The
solid line represents the saturation level ($L_\mathrm{X} / L_\mathrm{bol} = 10^{-3}$), while the
dotted line is for $L_\mathrm{X}/L_\mathrm{bol} = 10^{-4}$, and the dashed line for 
$L_\mathrm{X}/L_\mathrm{bol} = 10^{-5}$.\label{lxlb}}
\figcaption[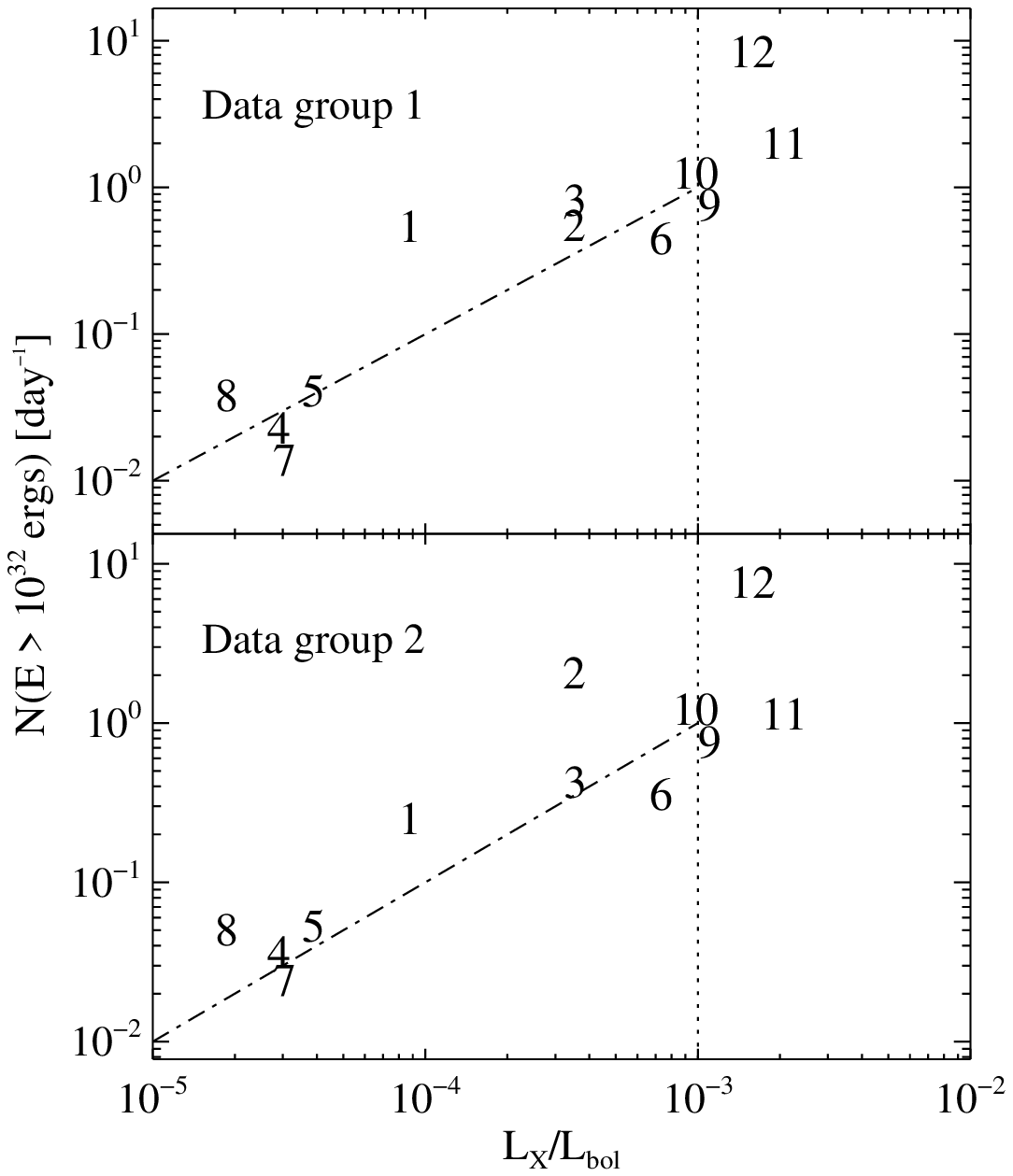]{Normalized occurrence rate of flares with energies larger than $10^{32}$~ergs
vs. ratio \lxlbol\  of the coronal luminosity and the bolometric
luminosity. Data groups 1 and 2 are identical to Fig.~\ref{distrlx}. The numbers are as in 
Fig.~\ref{lxlb}. The dash-dotted lines are lines with slope 1. The dotted
lines represent the saturation level ($L_\mathrm{X} = 10^{-3} L_\mathrm{bol}$). 
Note that, for clarity, the $N(>E_\mathrm{c})$ of points 4 and 9 have
been multiplied and divided by a factor of 1.5, respectively.\label{distrlxlb}}
\figcaption[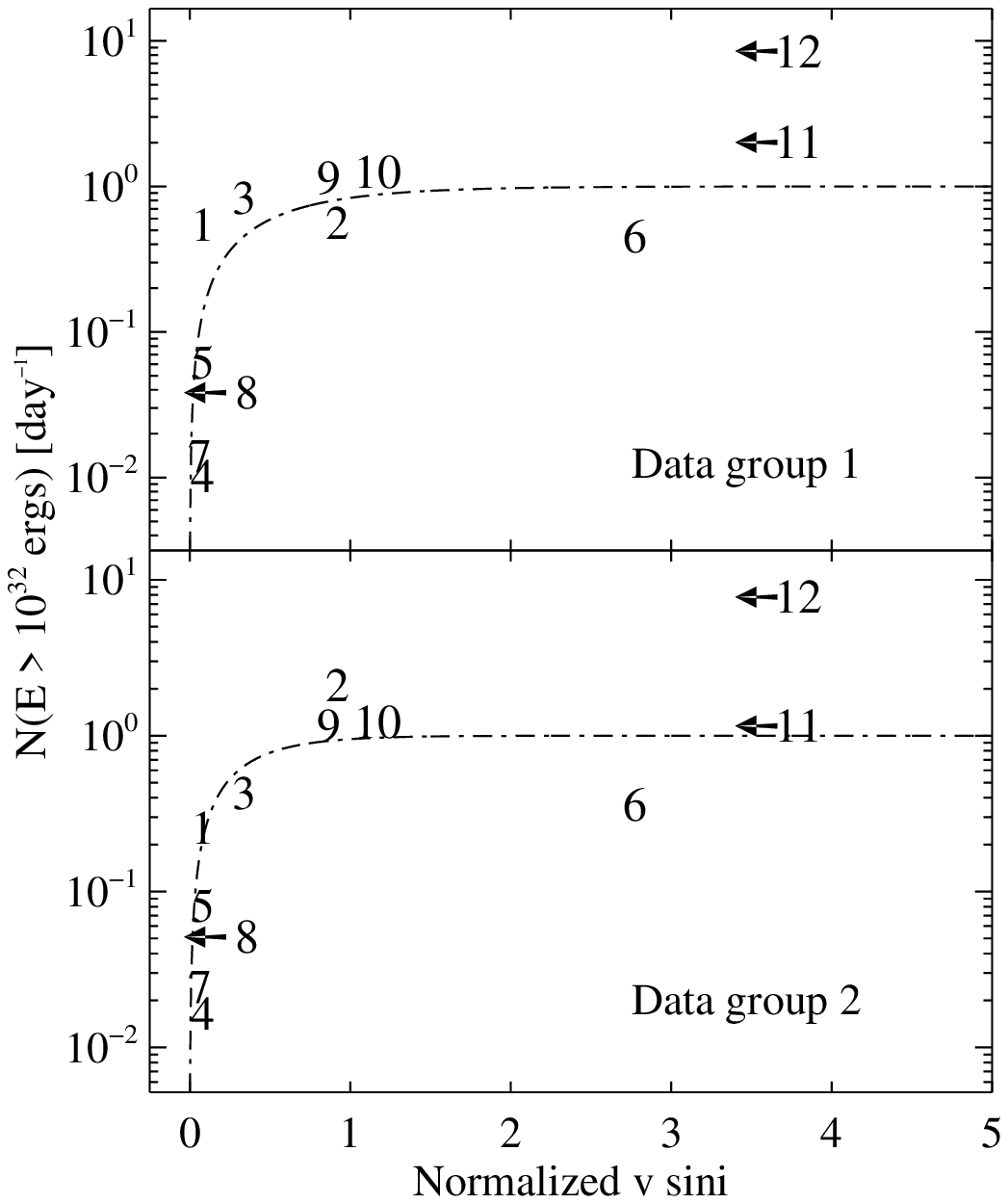]{Normalized occurrence rate of flares with energies larger 
than $10^{32}$~ergs vs. normalized projected
rotational velocity $v \sin i/(v \sin i)_\mathrm{sat} = x$ (spectral-type dependent). Data groups 1 and 2 are identical to Fig.~\ref{distrlx}. ``Saturation'' fits 
of the form $1-\exp (-x/{\zeta}$) are plotted dot-dashed. Arrows indicate upper
limits.  Note that, for clarity, the $N(>E_\mathrm{c})$ of points 4 and 5 have
been divided and multiplied by a factor of 1.5, respectively.\label{distrvsini}}
\figcaption[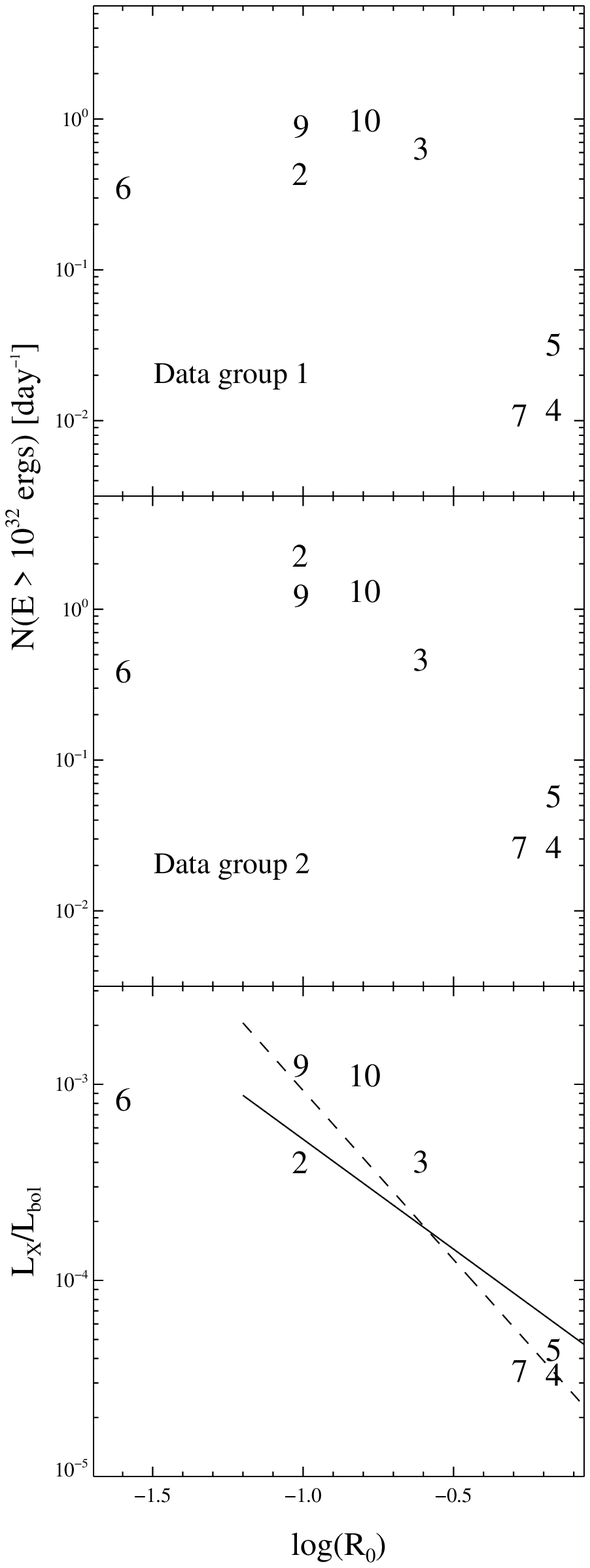]{(Upper and middle panels) Normalized flare rate vs.
Rossby number. Data groups and identifications are as in Fig.~\ref{lxlb}.
(Lower panel) Ratio \lxlbol\ vs. the Rossby number for our sample. The solid
line corresponds to the fit by \citet{rand96}, while the dashed line refers to our
best fit (see text).\label{distrross}}
\figcaption[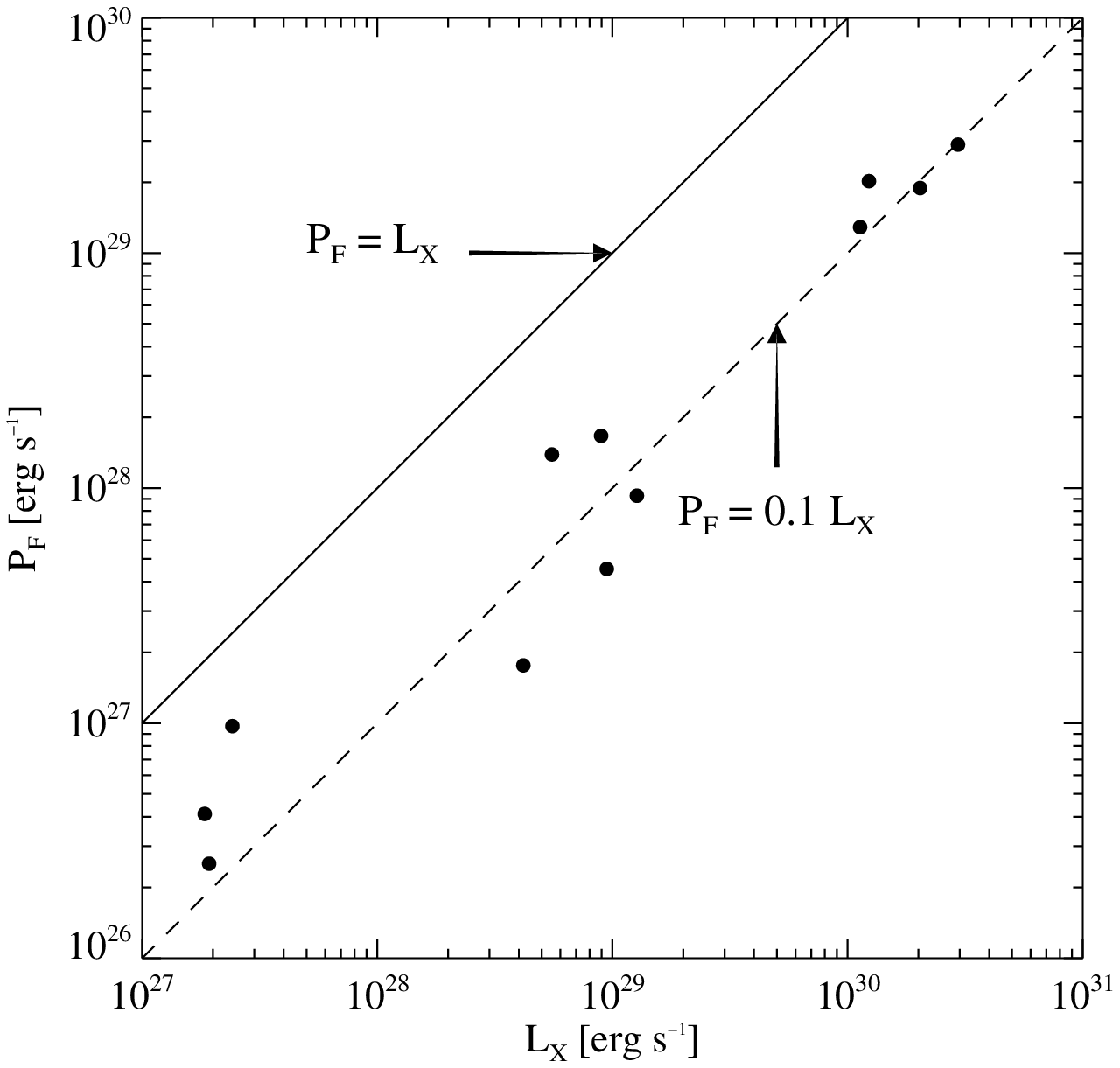]{X-radiated power $P_F$ from the detected flares
 vs. coronal luminosity $L_\mathrm{X}$. The
solid line represents proportionality ($P_F = L_\mathrm{X}$), while the
dashed line is for $P_F = 0.1 L_\mathrm{X}$.\label{en}}

\clearpage
\begin{deluxetable}{llcccccccccc@{$\; \rightarrow \; $}c}
\tabletypesize{\small}
\tablecaption{Target Selection List.\label{list}}
\tablewidth{0pt}
\tablecolumns{13}
\tablehead{
\colhead{Source} & 
\colhead{Spectral} & 
\colhead{$d$} &
\colhead{$P$} & 
\colhead{$\mathrm{Ref_1}$} & 
\colhead{$v \sin i$} & 
\colhead{$\mathrm{Ref_2}$} & 
\colhead{\bv} &
\colhead{$V$} &
\colhead{$\mu$} &
\colhead{$\log L_\mathrm{X}$} &
\multicolumn{2}{c}{\euve}\\
\colhead{Name} &
\colhead{Type} &
\colhead{(pc)} &
\colhead{(d)} &
\colhead{} &
\colhead{(\kms)} &
\colhead{} &
\colhead{(mag)} &
\colhead{(mag)} &
\colhead{(\cps)} &
\colhead{(\ergps)} &
\multicolumn{2}{c}{Observing Window}}
\startdata
\objectname[]{HD 2726}  	    & F2~V  & 45.07 & \nodata & \nodata & 13.2 & 1  & 0.367 & 5.67 & 0.11 & 30.47 & 1995/08/09     & 1995/08/16\\
\objectname[]{47 Cas}		    & G0--5~V& 33.56& 1.0   & 1 & 62.1\tablenotemark{a} & \nodata & 0.620\tablenotemark{a} & \nodata\tablenotemark{a} & 0.14 & 30.31 & 1997/01/23 &  1997/01/29\\
\objectname[]{EK Dra}		    & G1.5~V	    & 33.94 & 2.605 & 2     & 17.3  & 2     & 0.626 & 7.60 & 0.08 & 30.09 & 1995/12/06 & 1995/12/13\\
\objectname[]{$\kappa$ Cet}	    & G5~V  & 9.16  & 9.4   & 3     & 3.9   & 3     & 0.681 & 4.84 & 0.08 & 28.98 & 1994/10/13 & 1994/10/18\\
{}			    & {}    & {}    & {} & {} &     {}      & {}    & {} & {} & 0.11 & 29.10 & 1995/10/06  & 1995/10/13\\
\objectname[]{AB Dor}		    & K1~V  & 14.94 & 0.515 & 4 & 93.0  	    & 4 & 0.830 & 6.88 & 0.28 & 30.05 & 1994/11/12 & 1994/11/17\\
\objectname[]{$\epsilon$ Eri}	    & K2~V  & 3.22  & 11.3 & 5 &    2.0 	    & 3 & 0.881 & 3.72 & 0.30 & 28.62 & 1995/09/05 & 1995/09/13\\
\objectname[]{GJ 411}		    & M2~V  & 2.55  & \nodata & \nodata & $<2.9$ & 5 & 1.502 & 7.49 & 0.02 & 27.29 & 1995/03/22 & 1995/04/04\\
\objectname[]{AD Leo}		    & M3~V  & 4.90  & 2.7   & 6 & 6.2		    & 5 & 1.540 & 9.43 & 0.28 & 28.95 & 1996/05/03 & 1996/05/06\\
\objectname[]{EV Lac}		    & M4.5~V & 5.05 & 4.376 & 7 & 6.9		    & 5     & 1.540 & 10.29 & 0.16 & 28.74 & 1993/09/09 & 1993/09/13\\
\objectname[]{CN Leo}		    & M6~V & 2.39   & \nodata & \nodata & $<2.9$ & 5 & 2.000 & 13.54 & 0.03 & 27.38 & 1994/12/16 & 1994/12/19\\
{}			    & {}    & {}    & {} & {} & {}	    & {}    & {} & {} & 0.02 & 27.27 & 1995/01/24  & 1995/01/30
\enddata
\tablenotetext{a}{$v \sin i$: Estimated equatorial velocity; \bv: Value set to
account for the X-ray star's spectral type; $V$: no data for X-ray emitter; (see
text)}
\tablerefs{Rotation period ($\mathrm{Ref_1}$). (1) \citealt*{guedel95}, 
(2) \citealt*{strass97}, 
(3) \citealt{noy84}, 
(4) \citealt{innis88}, 
(5) \citealt{bal83}, 
(6) \citealt{spies86},
(7) \citealt*{pett92}.
Projected rotational velocity ($\mathrm{Ref_2}$).
(1) \citealt*{groot96}, 
(2) \citealt{strass98},  
(3) \citealt{fek97}, 
(4) \citealt*{kurst94}, 
(5) \citealt{delf98}}
\end{deluxetable}

\clearpage
\begin{deluxetable}{lcc@{~~(}c@{,~}c@{)~~~}ccc@{~~(}c@{,~}c@{)}ccc@{~~(}c@{,~}c@{)}}
\tablecaption{
Fits To The Flare Rate Distributions in Energy.
\label{tabresult}}
\tablewidth{0pt}
\tablecolumns{15}
\tablehead{
\colhead{} & 
\multicolumn{5}{c}{Cumulative} & 
\colhead{} & 
\multicolumn{3}{c}{Differential} &
\colhead{} &
\multicolumn{4}{c}{Simultaneous}\\
\cline{2-6} \cline{8-10} \cline{12-15}\\
\colhead{Name} & 
\colhead{$\alpha$\tablenotemark{a}} & 
\multicolumn{3}{c}{$\log(E_\mathrm{min})$\tablenotemark{d}~~[ergs]} &
\colhead{$\alpha$\tablenotemark{b}} & 
\colhead{} & 
\multicolumn{3}{c}{$\alpha$\tablenotemark{c}} &
\colhead{} &
\colhead{Type} &
\multicolumn{3}{c}{$\alpha$\tablenotemark{c}}
}
\startdata
HD 2726 	      & $2.61 \pm 0.38$ & 31.7 &  29.7 & 32.3 & 1.89\tablenotemark{e} & {} & 2.43 & 1.80 & 3.93 &{} & F+G & 2.28 & 2.03 & 2.57\\
47 Cas  	      & $2.19 \pm 0.34$ & 29.7 &  \nodata& 31.6 & 1.98 & {} & 2.62 & 1.72 & 5.41&{}&\vline&\multicolumn{3}{c}{}\\
EK Dra  	      & $2.08 \pm 0.34$ & 30.2 &  \nodata & 32.0 & 2.27 & {} & 1.78 & 1.26 & 2.49&{}&\vline&\multicolumn{3}{c}{}\\
$\kappa$ Cet 1994     & $2.18 \pm 0.89$ & 27.2 &  \nodata & 31.0 & 1.90 & {} & 2.55 & 0.31 & \nodata&{}&\vline&\multicolumn{3}{c}{}\\
$\kappa$ Cet 1995     & $2.29 \pm 0.51$ & 29.5 &  \nodata & 31.1 & 2.21 & {} & 2.45 & 1.65 & 3.80&{}&\vline&\multicolumn{3}{c}{}\\
AB Dor  	      & $1.88 \pm 0.26$ & \nodata  & \nodata & 28.8 & 1.97 & {} & 1.76 & 1.24 & 2.72 &{}& K & 1.87 & 1.50 & 2.39\\
$\epsilon$ Eri        & $2.40 \pm 0.81$ & 29.1 &  \nodata & 30.7 & 2.50 & {} & 2.38 & 1.06 & 4.05&{}&\vline&\multicolumn{3}{c}{}\\
GJ 411  	      & $1.63 \pm 0.29$ & \nodata & \nodata & \nodata & 1.96 & {} & 1.57 & 1.08 & 2.22 & &{}M & 1.84 & 1.63 & 2.06\\
AD Leo  	      & $2.02 \pm 0.28$ & 26.2 &  \nodata & 29.8 & 1.85 & {} & 1.65 & 1.18 & 2.35&{}&\vline&\multicolumn{3}{c}{}\\
EV Lac  	      & $1.76 \pm 0.33$ & \nodata  & \nodata & 29.1 & 1.90 & {} & 1.75 & 0.98 & 3.33&{}&\vline&\multicolumn{3}{c}{}\\
CN Leo 1994	      & $2.21 \pm 0.30$ & 29.3 &  27.0 & 29.8 & 1.91 & {} & 2.24 & 1.78 & 3.04&{}&\vline&\multicolumn{3}{c}{}\\
CN Leo 1995	      & $1.46 \pm 0.39$\tablenotemark{f} & \nodata & \nodata &\nodata& 2.14 & {} & 1.59 & 0.84 & 2.56&{}&\vline&\multicolumn{3}{c}{}
\enddata
\tablenotetext{a}{From an adapted version of \citet{craw70}}
\tablenotetext{b}{$\chi^2$ linear fit in the $\log - \log$ plane}
\tablenotetext{c}{$\chi^2$ fit within XSPEC with 68~\% confidence ranges for a single parameter}
\tablenotetext{d}{Minimum energy $E_\mathrm{min}$ required for the power law to explain the 
total observed radiative energy loss; limits are given in parentheses}
\tablenotetext{e}{Influenced by the largest flare energy; $\alpha = 2.43$ if
removed}
\tablenotetext{f}{Influenced by the flat low-energy end of the distribution}
\end{deluxetable}


\clearpage
\begin{deluxetable}{cr@{~(}c@{)~~~}r@{~(}c@{)~~~}r@{~(}c@{)~~~}r@{~(}c@{)~~~}r@{~(}c@{)}}
\tablecaption{Correlation tests for the power-law index $\alpha$.
\label{tabcorr}}
\tablewidth{0pt}
\tablecolumns{1}
\tablehead{
\colhead{$\alpha$ versus:} &
\multicolumn{2}{c}{$L_\mathrm{X}$} &
\multicolumn{2}{c}{\lxlbol} &
\multicolumn{2}{c}{$P$} &
\multicolumn{2}{c}{$v \sin i$} &
\multicolumn{2}{c}{$R_0$}
}
\startdata
\sidehead{Spearman test:\tablenotemark{a}}
DG1 & 0.50 & 0.10 & $-0.28$ & 0.38 & 0.50 & 0.25 & $-0.36$ & 0.38 & 0.54 & 0.22  \\
DG2 & 0.63 & 0.03 & $-0.34$ & 0.28 & 0.07 & 0.88 & 0.05 & 0.91 & 0.18 & 0.70\\
\sidehead{Kendall's $\tau$ test:\tablenotemark{a}}
DG1 & 0.42 & 0.05 & $-0.24$ & 0.27 & 0.33 & 0.29 & $-0.29$ & 0.32 & 0.43 & 0.17\\ 
DG2 & 0.45 & 0.04 & $-0.33$ & 0.13 & 0.05 & 0.88 & 0.07 & 0.80 & 0.14 & 0.65
\enddata
\tablenotetext{a}{Spearman's rank-correlation coefficients $r_\mathrm{S}$ and
Kendall's coefficients $\tau$. In parentheses, the two-sided significances
of their deviation from zero ($\equiv$no correlation)}
\end{deluxetable}



\onecolumn
\centering
\setcounter{figure}{0}
\clearpage
\begin{figure}
\epsscale{0.99}
\plotone{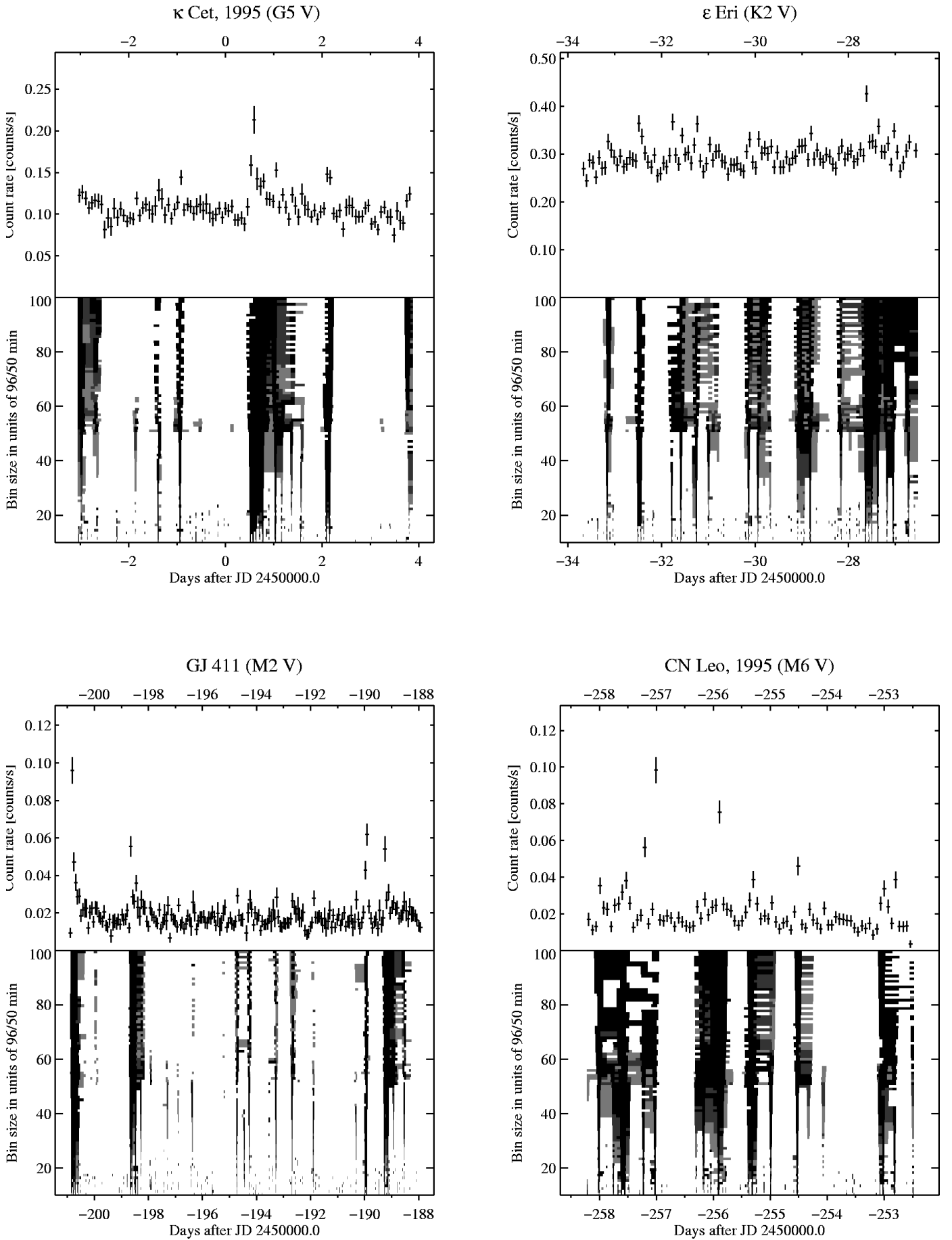}
\caption{Examples of light curves and significance plots.}
\end{figure}

\clearpage
\begin{figure}
\epsscale{0.99}
\plotone{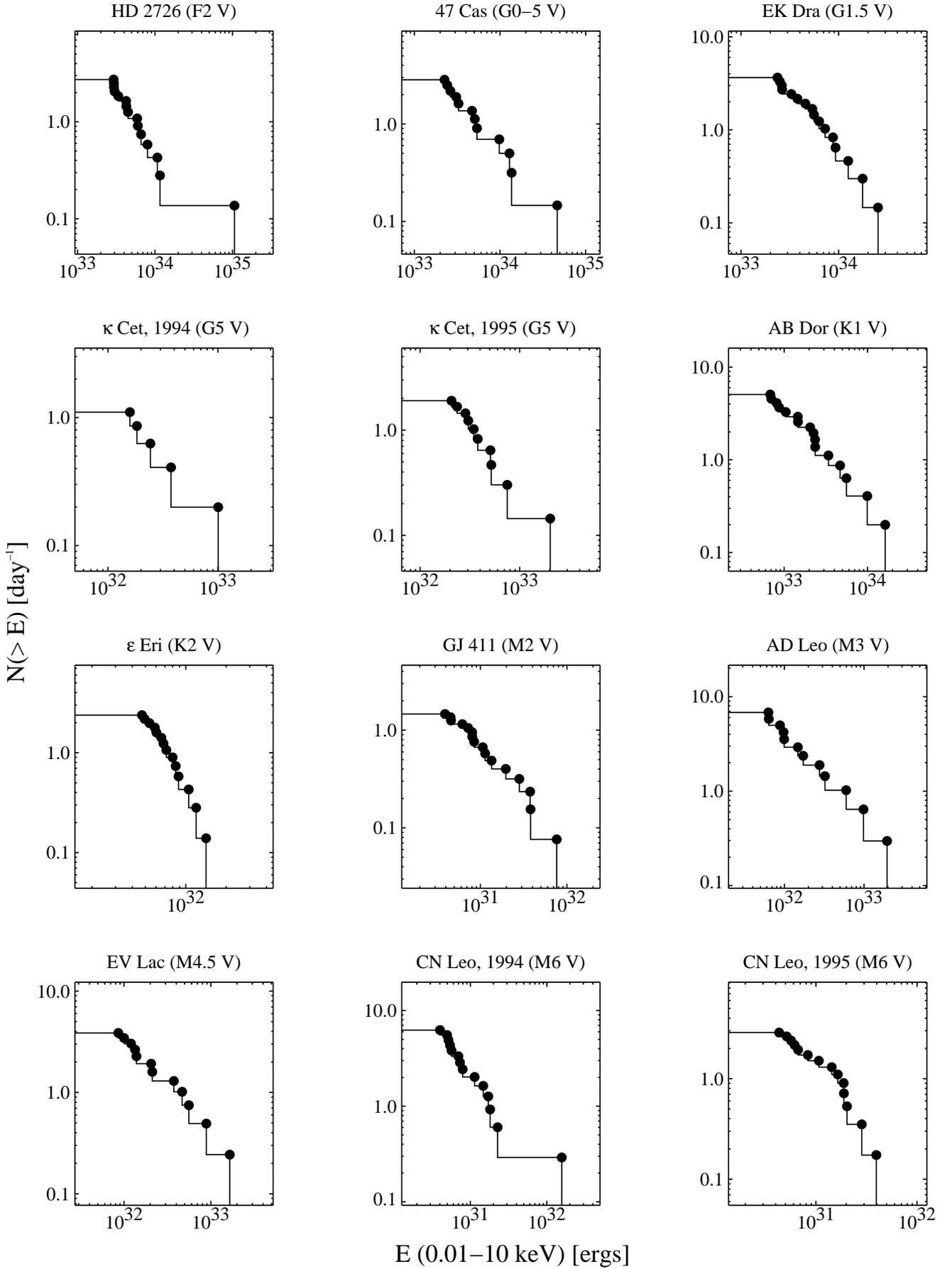}
\caption{Cumulative flare occurrence rate constructed for all twelve
data sets.}
\end{figure}

\clearpage
\begin{figure}
\epsscale{0.5}
\plotone{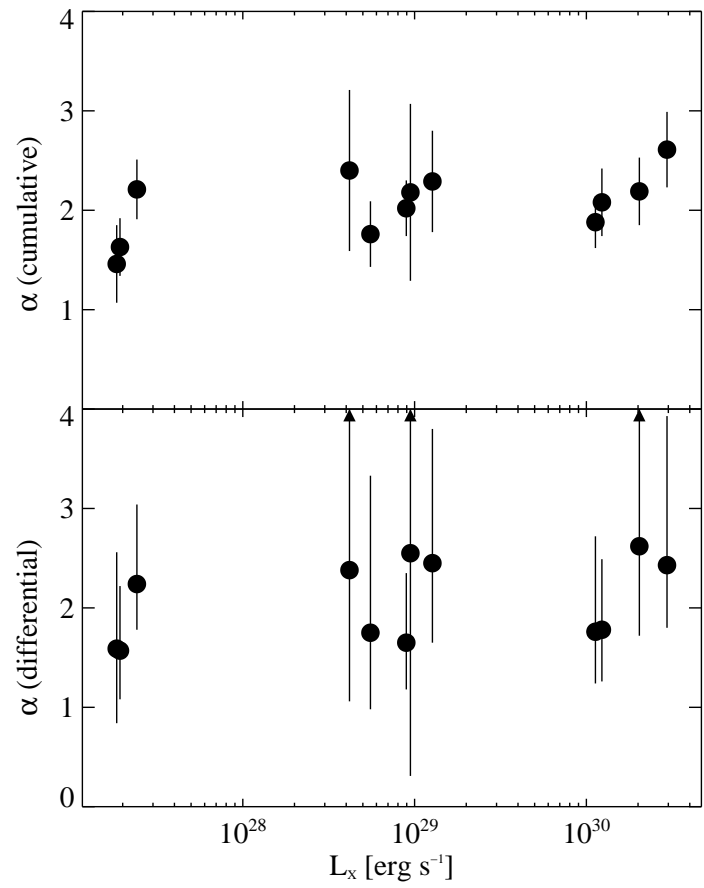}
\caption{Power-law index $\alpha$ vs. coronal luminosity $L_\mathrm{X}$.}
\end{figure}

\clearpage
\begin{figure}
\epsscale{0.5}
\plotone{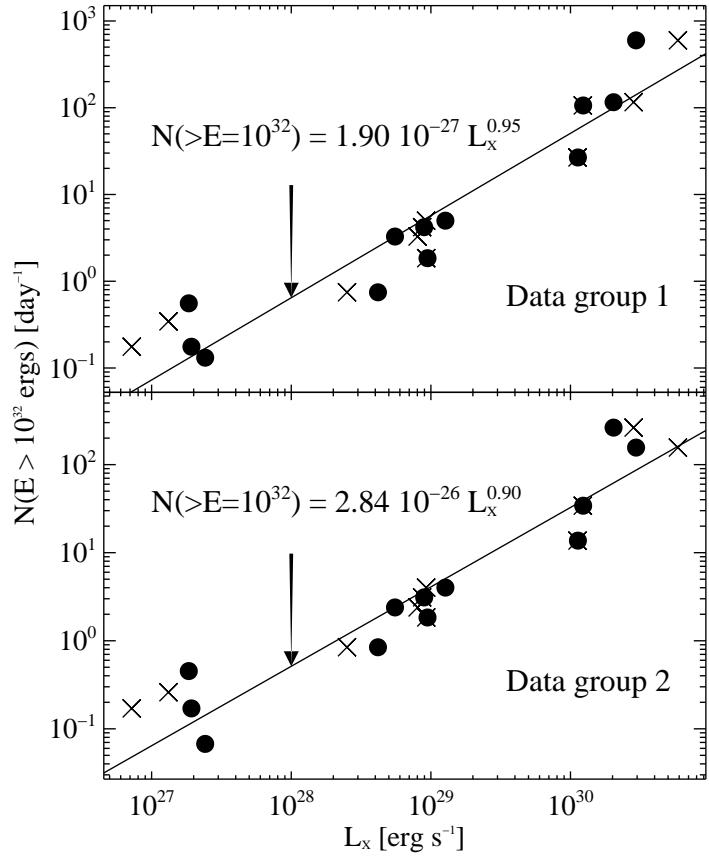}
\caption{Occurrence rates of flares with energies larger than $10^{32}$~ergs
vs. coronal luminosity $L_\mathrm{X}$. Data group 1 corresponds to the rates 
derived from the fits to cumulative distributions, while data group 2 corresponds to 
the rates derived from the fits to differential distributions. The $\chi^2$ linear best-fits
in the $\log - \log$ plane are shown as straight lines together with the analytical 
formulation. Crosses represent $L_\mathrm{X}$ values for these sources taken from the 
literature (see text).}
\end{figure}

\clearpage
\begin{figure}
\epsscale{1.0}
\plotone{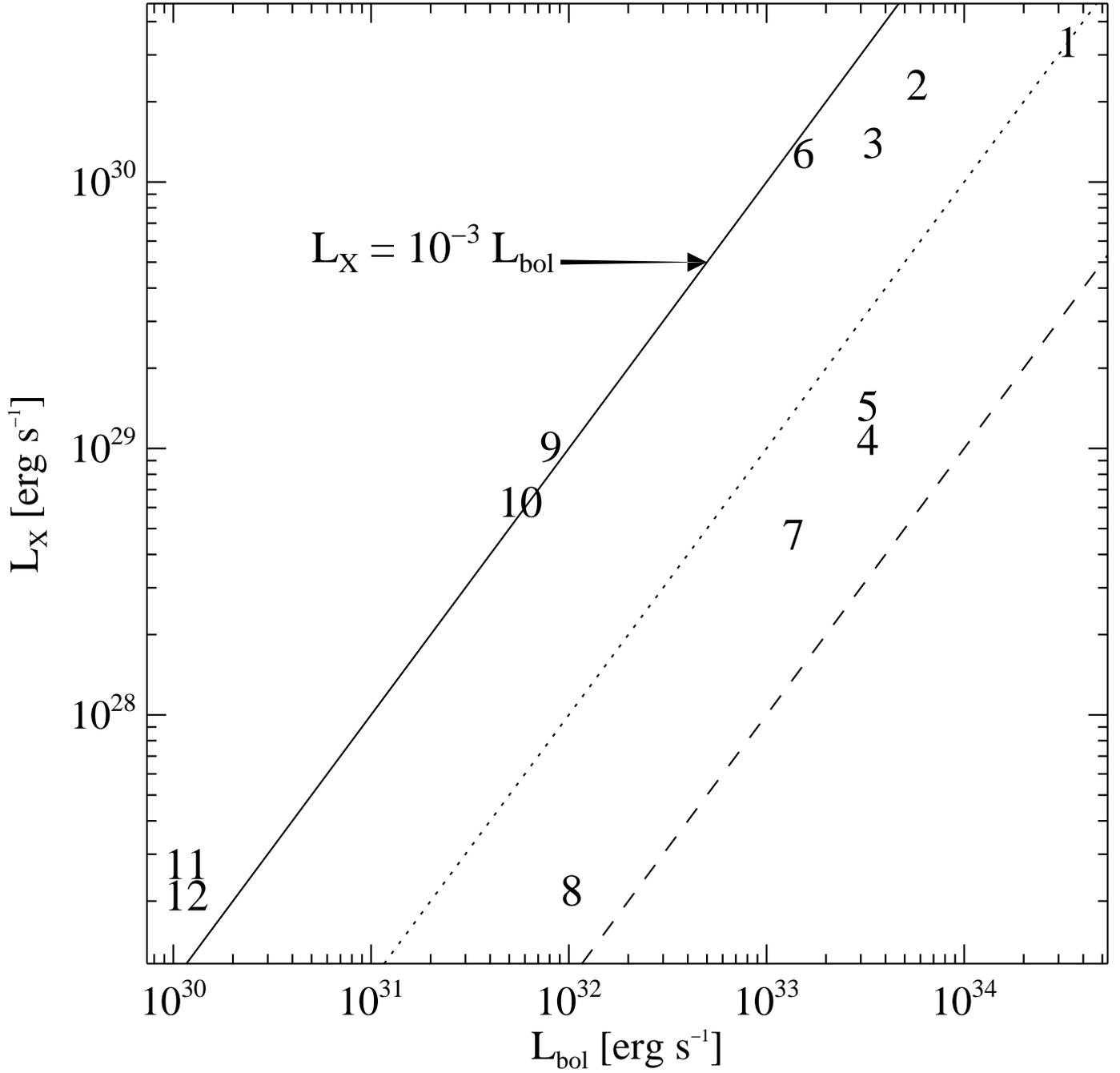}
\caption{Coronal luminosity $L_\mathrm{X}$ vs. bolometric luminosity
$L_\mathrm{bol}$. The source identification numbers refer to Table~\ref{list}. The
solid line represents the saturation level ($L_\mathrm{X} / L_\mathrm{bol} = 10^{-3}$), while the
dotted line is for $L_\mathrm{X}/L_\mathrm{bol} = 10^{-4}$, and the dashed line for 
$L_\mathrm{X}/L_\mathrm{bol} = 10^{-5}$.}
\end{figure}

\clearpage
\begin{figure}
\epsscale{0.5}
\plotone{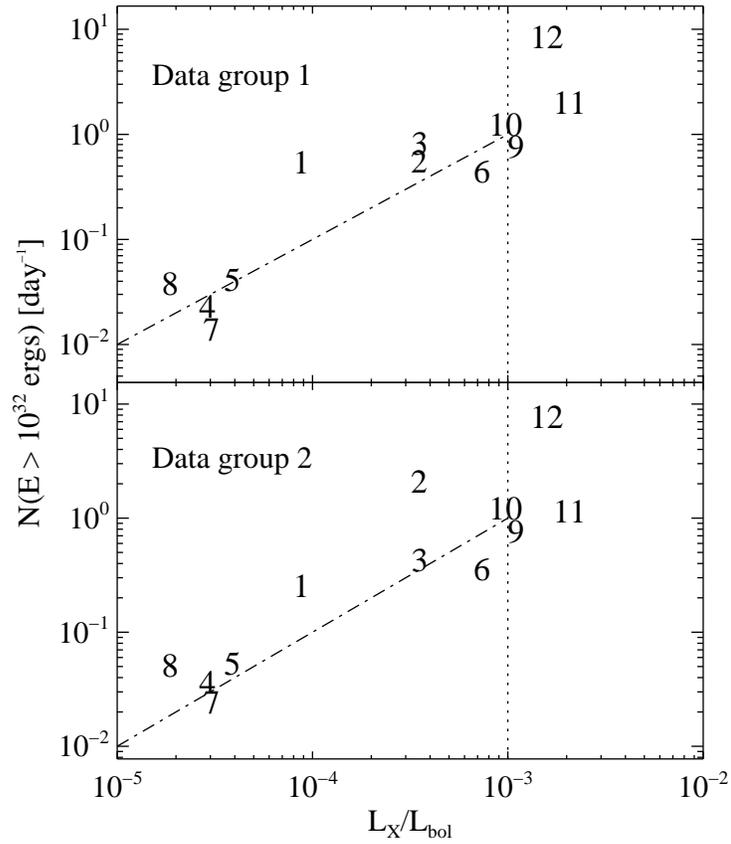}
\caption{Normalized occurrence rate of flares with energies larger than $10^{32}$~ergs
vs. ratio \lxlbol\  of the coronal luminosity and the bolometric
luminosity. Data groups 1 and 2 are identical to Fig.~\ref{distrlx}. The numbers are as in 
Fig.~\ref{lxlb}. The dash-dotted lines are lines with slope 1. The dotted
lines represent the saturation level ($L_\mathrm{X} = 10^{-3} L_\mathrm{bol}$).
Note that, for clarity, the $N(>E_\mathrm{c})$ of points 4 and 9 have
been multiplied and divided by a factor of 1.5, respectively.}
\end{figure}

\clearpage
\begin{figure}
\epsscale{0.5}
\plotone{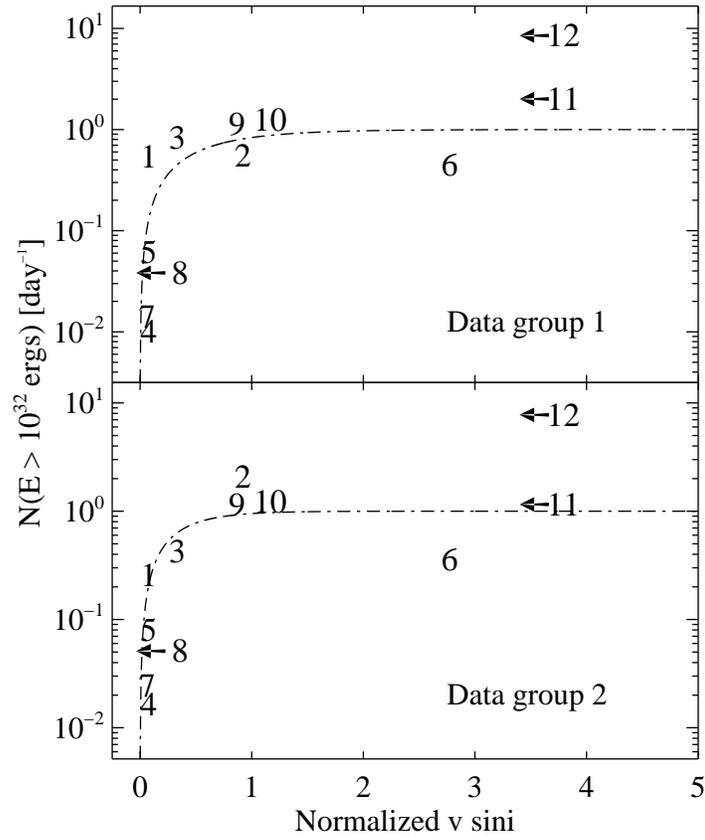}
\caption{Normalized occurrence rate of flares with energies larger 
than $10^{32}$~ergs vs. normalized projected
rotational velocity $v \sin i/(v \sin i)_\mathrm{sat} = x$ (spectral-type dependent). 
Data groups 1 and 2 are identical to Fig.~\ref{distrlx}. ``Saturation'' fits 
of the form $1-\exp (-x/{\zeta}$) are plotted dot-dashed. Arrows indicate upper
limits. Note that, for clarity, the $N(>E_\mathrm{c})$ of points 4 and 5 have
been divided and multiplied by a factor of 1.5, respectively.}
\end{figure}

\clearpage
\begin{figure}
\epsscale{0.47}
\plotone{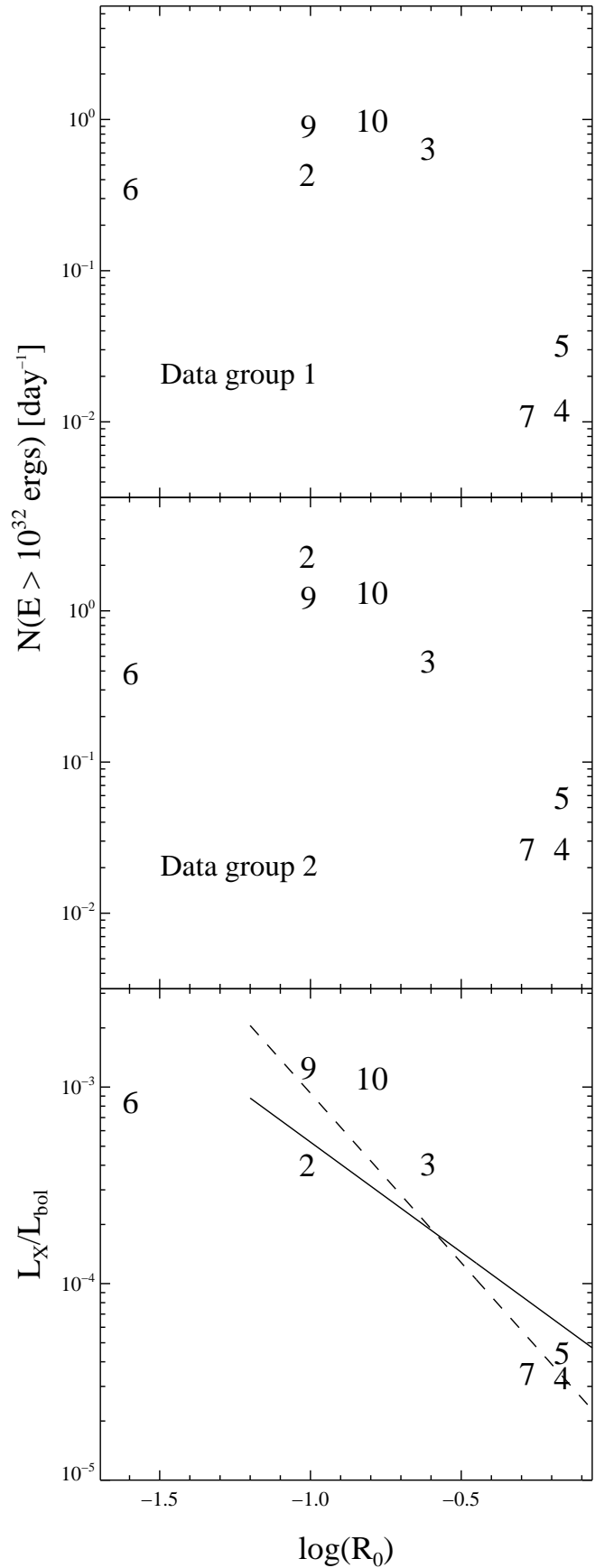}
\caption{(Upper and middle panels) Normalized flare rate vs.
Rossby number. Data groups and identifications are as in Fig.~\ref{lxlb}.
(Lower panel) Ratio \lxlbol\ vs. the Rossby number for our sample. The solid
line corresponds to the fit by \citet{rand96}, while the dashed line refers to our
best fit (see text).}
\end{figure}

\clearpage
\begin{figure}
\epsscale{1.0}
\plotone{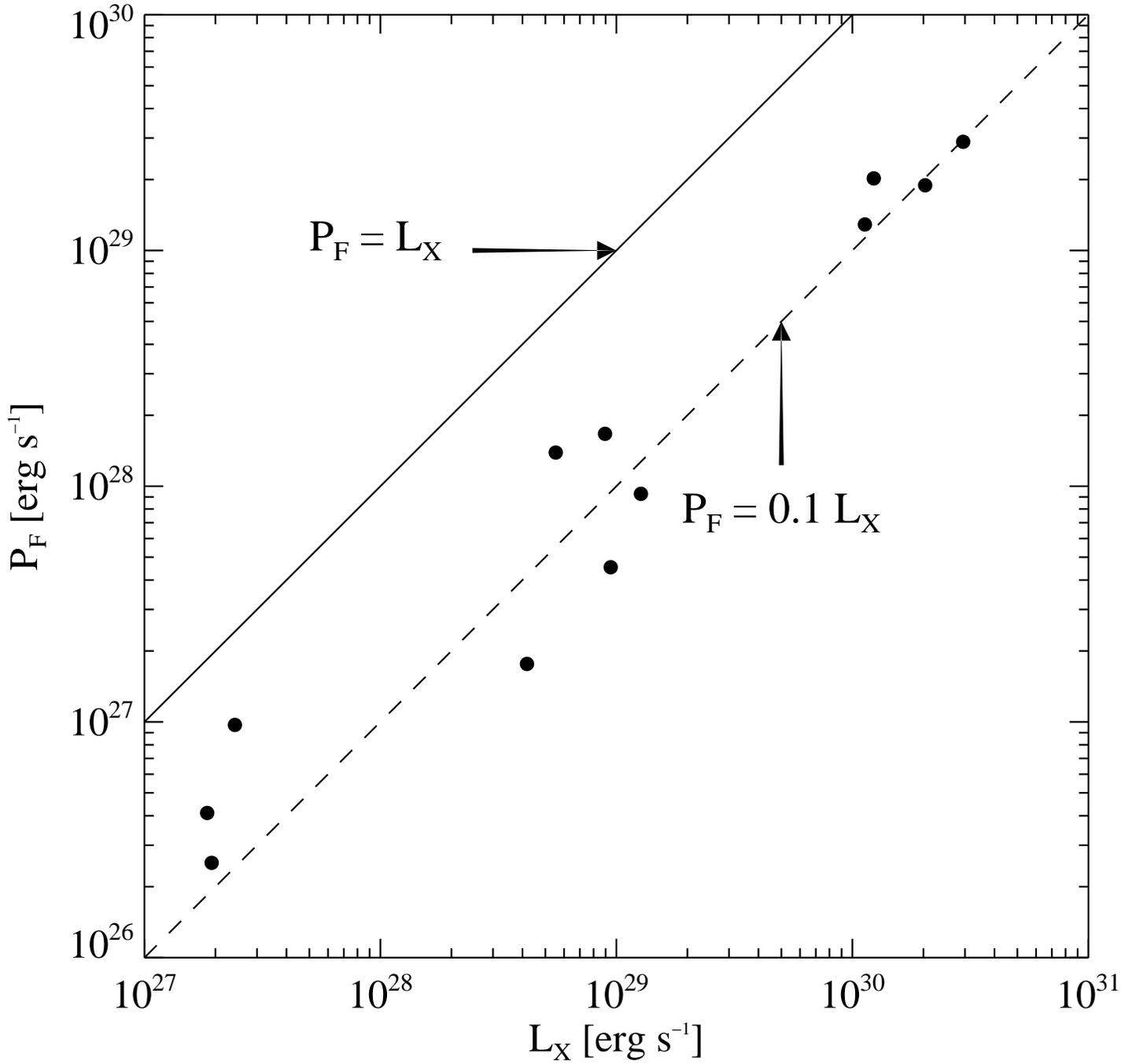}
\caption{X-radiated power $P_F$ from the detected flares
 vs. coronal luminosity $L_\mathrm{X}$. The
solid line represents proportionality ($P_F = L_\mathrm{X}$), while the
dashed line is for $P_F = 0.1 L_\mathrm{X}$.}
\end{figure}

\end{document}